\documentclass[aps,prx,twocolumn,showpacs,superscriptaddress]{revtex4-1}
\usepackage{amsmath,braket,amssymb,graphicx,hyperref,color,bbm,tabularx}
\usepackage[normalem]{ulem} %\sout for strikeout
\hypersetup{colorlinks, breaklinks, linkcolor={blue}, citecolor={blue}, urlcolor=blue}

\newcolumntype{Y}{>{\centering\arraybackslash}X}
\newcommand{\ignore}[1]{}

\begin{document}

\title{Topological phase transition and the effect of Hubbard interaction on the one-dimensional topological Kondo insulator}

\author{Jason C. Pillay}
\email{j.pillay@uq.edu.au}
\affiliation{ARC Centre for Engineered Quantum Systems, School of Mathematics and Physics, The University of Queensland, St Lucia, QLD 4072, Australia}

\author{Ian P. McCulloch}
\email{ianmcc@physics.uq.edu.au}
\affiliation{ARC Centre for Engineered Quantum Systems, School of Mathematics and Physics, The University of Queensland, St Lucia, QLD 4072, Australia}

\date{\today}

%%%%%%%%%%%%%%%%%%%%%%%%%%%%%%%%%%%%%%%%%%%%%%%%%%%%%%%%%%%%%%%%%%%%%%%%%%%%%%%%%%%%%%%%%%%%%%%%%%%%%%%%%%%
%%%%%%%%%%%%%%%%%%%%%%%%%%%%%%%%%%%%%%%%%%%%%%%%%%%%%%%%%%%%%%%%%%%%%%%%%%%%%%%%%%%%%%%%%%%%%%%%%%%%%%%%%%%

\begin{abstract}
The effect of a local Kondo coupling and Hubbard interaction on the topological phase of the one-dimensional topological Kondo insulator (TKI) is numerically investigated using the infinite matrix-product state density-matrix renormalization group algorithm. The groundstate of the TKI is a symmetry-protected topological (SPT) phase protected by inversion symmetry. It is found that on its own, the Hubbard interaction that tends to force fermions into a one-charge per site order is insufficient to destroy the SPT phase. However when the local Kondo Hamiltonian term that favors a topologically trivial groundstate with a one-charge per site order is introduced, the Hubbard interaction assists in the destruction of the SPT phase. This topological phase transition occurs in the charge sector where the correlation length of the charge excitation diverges while the correlation length of the spin excitation remains finite. The critical exponents, central charge and the phase diagram separating the SPT phase from the topologically trivial phase are presented.
\end{abstract}

\maketitle

%%%%%%%%%%%%%%%%%%%%%%%%%%%%%%%%%%%%%%%%%%%%%%%%%%%%%%%%%%%%%%%%%%%%%%%%%%%%%%%%%%%%%%%%%%%%%%%%%%%%%%%%%%%
%%%%%%%%%%%%%%%%%%%%%%%%%%%%%%%%%%%%%%%%%%%%%%%%%%%%%%%%%%%%%%%%%%%%%%%%%%%%%%%%%%%%%%%%%%%%%%%%%%%%%%%%%%%

\section{Introduction}
Theoretical studies of the low-temperature resistivity plateau in SmB$_6$ has brought rich, new physics in describing how strong spin-orbit coupling can give rise to non-trivial topological phases with exotic physical manifestations such as robust metallic surface states \cite{Dzero, Dzero2, Alexandrov}. In these studies, the three-dimensional SmB$_6$ was modeled as a Kondo insulator with spin-orbit coupling that hybridized conduction and $f$-electrons. Known as the topological Kondo insulator (TKI), the main factor that contributes to the non-trivial topological property of this model is the odd parity of the $f$-electron orbital structure that respects spatial-inversion and time-reversal symmetry \cite{Dzero}.

From these phenomenological studies, a one-dimensional (1D) TKI model was proposed in Ref. [\onlinecite{Alexandrov2}] to further gain insight into the properties of such interacting topological insulators. The non-trivial topology in this 1D model was realized by using a non-local coupling between an electron and its neighbouring local moment which mimics the large momentum $f$-electron orbital in the three-dimensional TKI. In the non-interacting limit, mean-field calculations showed this non-local coupling formed odd parity bands that invert under hybridization, thus forming an interacting topological band insulator \cite{Alexandrov2}. With weak interactions, bosonization and renormalization-group techniques have shown that the groundstate of the 1D TKI behaves as a spin-1 Haldane chain classified by a $Z_2$ topological invariant, and possesses spin-$\frac{1}{2}$ magnetic end states \cite{Lobos}.

Mean-field treatments and bosonization method fail when strong interactions are to be taken into account. This is where numerical methods come in. Being a heavy-weight in 1D simulations, the density-matrix renormalization group (DMRG) algorithm has shown to be a powerful tool in understanding groundstate properties of the 1D TKI model \cite{Mezio,Hagymasi,Lisandrini}. Current DMRG results show the existence of topologically protected spin-$\frac{1}{2}$ end-states and that the groundstate is in the Haldane phase via a string-order parameter \cite{Mezio}, thus confirming the topological origin of the end-states. Besides this, the stability of the Haldane phase against Ising anisotropy and Hubbard interaction was studied in Ref. \cite{Hagymasi} where it was found that the Hubbard interaction induced a phase transition into a N\'{e}el state only when the Ising anisotropy was non-zero. Another interesting study \cite{Lisandrini} showed that when the conventional local $s$-wave Kondo coupling was introduced together with the non-local $p$-wave Kondo coupling, a topological phase transition occurred when the former's coupling constant exceeded a critical value.

So far, these numerical works have been limited to finite-length lattices with open boundaries, thus thermodynamic properties were obtained through finite-size scaling of the system's size. In addition to that, the basis size (or bond dimension) of the wavefunction also puts a limit on the exact representation of the wavefunction and thus the accuracy of the numerical data. To overcome this, a second scaling is typically done in DMRG simulations - the scaling of data with respect to bond dimension. In this work, the infinite matrix-product state DMRG \cite{McCulloch} is utilized and this gives direct access to the thermodynamic limit without having to carry out finite-size scaling of the lattice. Thus, the only scaling required is scaling of data with respect to the bond dimension.

The classification of symmetry-protected topological (SPT) phases is well understood in the sense that the possible SPT phases protected by the global symmetry group $G$ are given by the second group cohomology $\mathcal{H}^2(G, U(1))$ \cite{Chen,Chen1,Chen2}. However, not all global symmetries can protect SPT phases in a given physical system. In the Haldane phase of a spin-1 chain, the relevant global symmetries are spatial inversion ($\mathcal{I}$), time-reversal ($\mathcal{T}$) and dihedral ($\mathcal{D}_2$, the dihedral group of $\pi$-rotations about two orthogonal axes), which all separately protect the topological phase. On the other hand, in fermionic systems the presence of charge fluctuations reduce the possible protecting symmetries \cite{Fidkowski, Turner}. In numerical studies of interacting fermions \cite{Moudgalya, Nourse} it was shown that it is essential that an on-site representation of the symmetry is well-defined in order to protect the Haldane phase. When charge fluctuations are present, the on-site representations of $\mathcal{T}$ and $\mathcal{D}_2$ are ``graded" i.e. the respective symmetry representations are split into two separate representations that act separately on states containing even and odd number of fermions. The reason behind this is that the symmetry representations of $\mathcal{T}$ and $\mathcal{D}_2$ depend on the spin-1/2 fermion rotation operators $R^\alpha = \text{exp} \left( i \pi S^\alpha \right)$ ($\alpha = x, y, z$) whose product depend on the even or odd number of fermions at each site. When an even (odd) number of fermions are present at a site, the product of the representations of $R^\alpha$'s is even (odd). As a result, the product of the representations commute (anticommute) and its representation is a linear (projective) one.  Thus, the only symmetry that protects the Haldane phase is $\mathcal{I}$. This effect is most transparent in the entanglement spectrum (ES) and the ``non-local" order parameters
\begin{eqnarray}
O_\mathcal{I} &=& \braket{U_\mathcal{I} U_\mathcal{I}^*} \nonumber,
\\
O_\mathcal{T} &=& \braket{U_\mathcal{T} U_\mathcal{T}^*} \nonumber,
\\
O_{\mathcal{D}_2} &=& \braket{U_x U_z U_x^\dagger U_z^\dagger}.
\label{eqn:non-local-ord-param}
\end{eqnarray}
In the former, the grading of the integer (even number of fermions) and half-integer (odd number of fermions) representations give rise to a separation of a single pair of two-fold degenerate low-lying ES values into two pairs of two-fold degenerate low-lying ES values where each pair corresponds to the integer and half-integer states respectively. As for the non-local order parameters, the representation of the rotation operators are block-diagonalized into integer and half-integer parts when charge fluctuations are absent. Thus the matrix of the non-local operator (e.g. $U_x U_z U_x^\dagger U_z^\dagger$) contains $\pm 1$ along its diagonal depending whether the corresponding basis state transforms linearly or projectively and $|O_g| = 1$ ($g = \mathcal{I}, \mathcal{T}, \mathcal{D}_2$). When charge fluctuation is present, the grading of the representation causes both $-1$ and $+1$ sectors to gain contributions of non-zero Schmidt value from the groundstate and are thus simultaneously present in the diagonal matrix of the non-local operator matrix. As a result, the non-local order parameter $O_g$ of the symmetry whose representation is graded has a magnitude that is strictly less than 1, and it is not able to distinguish a topologically non-trivial phase from a topologically trivial one. Though a graded symmetry does not protect the Haldane phase \cite{Moudgalya, Nourse}, this is not true in general. This can be seen for instance in a topological superconductor where a graded time-reversal symmetry is fractionalized in the edge states in the topologically non-trivial phase \cite{Tang}.

The objective of this work is to explore in detail and classify the SPT phase of the 1D TKI toy model using an infinite matrix-product state (iMPS) DMRG approach. This approach is used to first study the effect of Hubbard interaction on the TKI groundstate. Second, the effect of the local $s$-wave Kondo coupling on the TKI groundstate is investigated and the topological phase transition between an SPT phase and topologically trivial phase is characterized. Finally, it is shown how the Hubbard interaction affects this topological phase transition.

%%%%%%%%%%%%%%%%%%%%%%%%%%%%%%%%%%%%%%%%%%%%%%%%%%%%%%%%%%%%%%%%%%%%%%%%%%%%%%%%%%%%%%%%%%%%%%%%%%%%%%%%%%%
%%%%%%%%%%%%%%%%%%%%%%%%%%%%%%%%%%%%%%%%%%%%%%%%%%%%%%%%%%%%%%%%%%%%%%%%%%%%%%%%%%%%%%%%%%%%%%%%%%%%%%%%%%%

\section{Model and Simulation Method}
\begin{figure}[h!]
  \centering\includegraphics[width=0.46\textwidth]{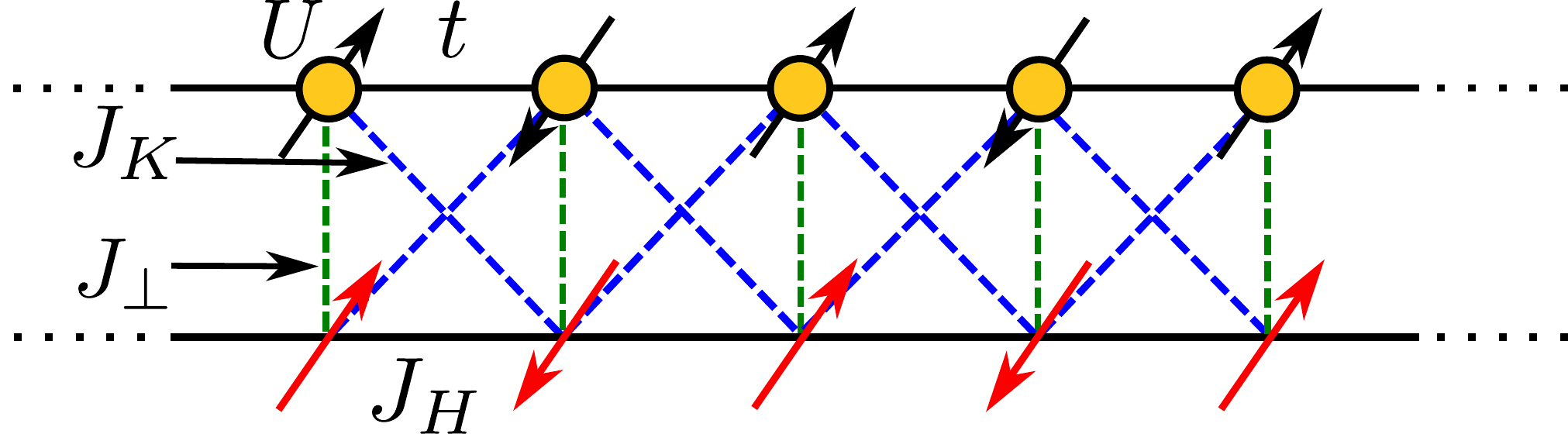}
  \caption{(Colour online) Schematic representation of a segment of the infinite 1D \textit{p}-wave Kondo-Heisenberg lattice and groundsate wavefunction. The top chain is a Hubbard chain and the bottom chain is an $S = \frac{1}{2}$ antiferromagnetic Heisenberg chain. The non-local Kondo exchange $J_K$ is a nearest-neighbour antiferromagnetic interaction ($J_K > 0$) that couples a spin $S_j$ at site $j$ in the Heisenberg chain to the \textit{p}-wave spin density $\pi_j$ in the Hubbard chain.}
  \label{fig:lattice}
\end{figure}

Fig. \ref{fig:lattice} depicts a segment of the infinite 1D topological Kondo insulator first proposed in Ref. \cite{Alexandrov2} which is also sometimes referred to as the $p$-wave Kondo-Heisenberg model. The Hamiltonian of this model is given by
\begin{eqnarray}
H = H_c + H_H + H_K + H_\perp,
\label{eqn_Hamiltonian_tot}
\end{eqnarray}
where
\begin{eqnarray}
H_c &=& -t \sum_{j,\sigma} \left( c^\dagger_{j+1, \sigma} c_{j, \sigma} + c^\dagger_{j, \sigma} c_{j+1, \sigma} \right) \nonumber
\\
&& + U \sum_j n_{j, \uparrow} n_{j, \downarrow}
\label{eqn_Hamiltonian_Hubbard}
\end{eqnarray}
is the 1D Hubbard Hamiltonian (top chain in Fig. \ref{fig:lattice}) describing fermions hopping with amplitude $t$ between sites $j$ and $j+1$, and a Hubbard interaction of strength $U$ between fermions of opposite spins at site $j$. The second term
\begin{eqnarray}
H_H = J_H \sum_j \vec{S}_j \cdot \vec{S}_{j+1} ,
\label{eqn_Hamiltonian_Heisenberg}
\end{eqnarray}
is the 1D Heisenberg Hamiltonian (bottom chain in Fig. \ref{fig:lattice}) describing spin exchange between nearest-neighbour $S = \frac{1}{2}$ localized spins. The third term $H_K$ represents the \textit{p}-wave Kondo coupling between the Hubbard and Heisenberg chains via a non-local Kondo exchange between electronic and spin degrees of freedom:
\begin{eqnarray}
H_K = J_K \sum_j \left[ \frac{1}{2} \left( S^+_j \pi^-_j + S^-_j \pi^+_j \right) + S^z_j \pi^z_j \right] .
\label{eqn_Hamiltonian_Kondo}
\end{eqnarray}
$S^\pm_j$ and $\pi^\pm_j$ ($S^z_j$ and $\pi^z_j$) are the ladder operators ($z$ components) of the spin $\vec{S}_j$ in the Heisenberg chain and the \textit{p}-wave spin density $\vec{\pi}_j$ in the Hubbard chain. The latter is given as
\begin{eqnarray}
\vec{\pi}_j = \frac{1}{2} \sum_{\alpha, \beta} p^\dagger_{j,\alpha} \vec{\sigma}_{\alpha,\beta} p_{j,\beta},
\label{eqn_pi_operator}
\end{eqnarray}
where $\vec{\sigma}$ is the vector of Pauli matrices, and
\begin{eqnarray}
p_{j,\sigma} = \frac{1}{\sqrt{2}} \left( c_{j+1,\sigma} - c_{j-1,\sigma} \right).
\label{eqn_p_operator}
\end{eqnarray}
The last term in Eq. (\ref{eqn_Hamiltonian_tot}) is the \textit{s}-wave coupling given by
\begin{eqnarray}
H_\perp = J_\perp \sum_j \left[ \frac{1}{2} \left( S^+_j s^-_j + S^-_j s^+_j \right) + S^z_j s^z_j \right],
\label{eqn_Hamiltonian_Kondo_local}
\end{eqnarray}
which describes the local exchange between the spin degrees of freedom of a fermion at site $j$ on the Hubbard chain with a localized spin at site $j$ on the Heisenberg chain.

The Hamiltonian Eq. (\ref{eqn_Hamiltonian_tot}) is symmetric under spatial inversion ($\mathcal{I}$)
\begin{eqnarray}
S^{x,y,z}_j \rightarrow S^{x,y,z}_{-j+1} \,\, , \,\, c_{j,\sigma} \rightarrow c_{-j+1,\sigma},
\label{eqn:inv_sym}
\end{eqnarray}
time-reversal symmetry ($\mathcal{T}$)
\begin{eqnarray}
S^{x,y,z}_j \rightarrow -S^{x,y,z}_j \,\, , \,\, c_{j,\uparrow} \rightarrow c_{j,\downarrow} \,\, , \,\, c_{j,\downarrow} \rightarrow -c_{j,\uparrow}
\label{eqn:time_rev_sym}
\end{eqnarray}
and $SU(2)$ (of which the dihedral group $\mathcal{D}_2$ is a subgroup).

The simulation is done using the infinite matrix-product state (iMPS) ansatz \cite{McCulloch, Schollwock}, where
\begin{eqnarray}
\ket{\psi} = \sum_{\left\lbrace j \right\rbrace } \left[ \ldots \Gamma_{j_1} \Lambda \Gamma_{j_2} \Lambda \ldots \right] \ket{\ldots j_1 j_2 \ldots}
\label{eqn:imps}
\end{eqnarray}
represents the wavefunction of the translationally invariant infinite lattice. Here $\Gamma_{j_i}$ is a $d \times m \times m$ tensor, where $d$ is the dimension of the local Hilbert space at site $i$ and $m$ is the basis size. $\Lambda$ is a $m \times m$ diagonal matrix that contains the Schmidt values of a bond between neighbouring sites and $j_i$ are the local degree of freedom at site $i$. Starting with a random iMPS, the wavefunction is variationally optimized using the infinite-DMRG (iDMRG) algorithm with single-site optimization \cite{McCulloch, Schollwock, Hubig} with a basis size ranging from $m = 900$ to 1500. This gives a groundstate wavefunction with a variance per site of the order of $10^{-6} - 10^{-11}$ and a truncation error of the order of $10^{-8} - 10^{-14}$. Utilizing SU(2) symmetry, the the basis size used here is approximately equivalent to $m = 2700 - 4500$ states of a U(1)-symmetric basis.

All data presented here are for $m \rightarrow \infty$, i.e. data scaled with respect to bond dimension $m$. This is done by collecting the relevant data at increments of $m$, and then scaling the data with respect to $m$ to obtain the value of the data at $m \rightarrow \infty$ \cite{Nishino,Andersson,Tagliacozzo,Crosswhite,McCulloch}. See the Appendix (Section \ref{appendix_scaling}) for more details of this procedure.

%%%%%%%%%%%%%%%%%%%%%%%%%%%%%%%%%%%%%%%%%%%%%%%%%%%%%%%%%%%%%%%%%%%%%%%%%%%%%%%%%%%%%%%%%%%%%%%%%%%%%%%%%%%
%%%%%%%%%%%%%%%%%%%%%%%%%%%%%%%%%%%%%%%%%%%%%%%%%%%%%%%%%%%%%%%%%%%%%%%%%%%%%%%%%%%%%%%%%%%%%%%%%%%%%%%%%%%

\section{Review of symmetries in iMPS and SPT order in 1D}
In this section, symmetries in the iMPS representation and the effects on the SPT phase from Ref. \cite{Pollman} is briefly reviewed. Attention is paid to $\mathcal{I}$, since $\mathcal{T}$ and $\mathcal{D}_2$ are graded and hence do not protect the Haldane phase. The $\Gamma$ and $\Lambda$ matrices of an iMPS satisfy the canonical condition
\begin{eqnarray}
\sum_j \Gamma_j^\dagger \Lambda^2 \Gamma_j = \mathbbm{1}
\label{eqn:imps2}
\end{eqnarray}
which can be understood as the transfer matrix
\begin{eqnarray}
T_{\alpha \alpha' ; \beta	\beta'} = \sum_j \Gamma^{\alpha}_{j \beta} \left( \Gamma^{\alpha'}_{j \beta'} \right)^* \Lambda_\beta \Lambda_{\beta'}
\label{eqn:transfer-matrix}
\end{eqnarray}
having a right eigenvector $\delta_{\beta \beta'}$ with eigenvalue 1, and
\begin{eqnarray}
\tilde{T}_{\alpha \alpha' ; \beta	\beta'} = \sum_j \left( \Gamma^{\alpha'}_{j \beta'} \right)^* \Gamma^{\alpha}_{j \beta} \Lambda_\alpha \Lambda_{\alpha'}
\label{eqn:transfer-matrix2}
\end{eqnarray}
having a left eigenvector $\delta_{\alpha \alpha'}$ also with eigenvalue 1.

The ES is the eigenvalues of the reduced density matrix \cite{Li} which is obtained by carrying out a Schmidt decomposition on the iMPS:
\begin{eqnarray}
\ket{\Psi} = \sum_\alpha \lambda_\alpha \ket{\alpha}_L \otimes \ket{\alpha}_R, \quad \ket{\alpha}_{L (R)} \in \mathcal{H}_{L (R)},
\label{eqn:schmidt-decomposition}
\end{eqnarray}
where $\lambda_\alpha$ are the Schmidt values and $\ket{\alpha}_{L (R)}$ are orthonormal basis states of the left (right) Hilbert space of the partition of the system. If the canonical condition Eq. (\ref{eqn:imps2}) is satisfied, $\lambda_\alpha$ are equal to the matrix elements $\Lambda_\alpha$. The set of eigenvalues of the reduced density matrix is $\lambda_\alpha^2$, which is also referred to as the entanglement spectrum. The entanglement entropy is defined as $S = \sum_\alpha \lambda_\alpha^2 \ln \lambda_\alpha^2$ and it corresponds to the von-Neumann entropy of the reduced density matrix. In a topologically trivial phase, there is no even or odd constraint to the degeneracy of $\Lambda_\alpha$. However in an SPT phase, the symmetry responsible for protecting the topological phase constrains $\Lambda_\alpha$ to be even-fold degenerate as will be shown below.

An iMPS that is invariant under local symmetry $g \in G$, which is represented in the spin basis as the unitary matrix $u_g$, satisfies
\begin{eqnarray}
\sum_{j j'} \left( u_g \right)_{j j'} \Gamma_{j'} = e^{i \theta_g} U^\dagger_g \Gamma_j U_g
\label{eqn:mps-transform}
\end{eqnarray}
where $U_g$ is a unitary matrix that commutes with the $\Lambda$ matrices and $e^{i\theta_g}$ is a phase factor. The left-hand side of Eq. (\ref{eqn:mps-transform}) varies for the different symmetries. For inversion, $u_{\mathcal{I}} = (-1)^{n_\uparrow n_\downarrow} \mathbbm{1}$ and $\Gamma_{j'}$ is replaced by $\Gamma_{j'}^T$ (transpose). The prefactor $(-1)^{n_\uparrow n_\downarrow}$ in $u_\mathcal{I}$ gives a $-1$ when inverting the doubly-occupied state $\ket{\uparrow \downarrow} \rightarrow \ket{\downarrow \uparrow} = -\ket{\uparrow \downarrow}$ or $\ket{\downarrow \uparrow} \rightarrow \ket{\uparrow \downarrow} = -\ket{\downarrow \uparrow}$, but leaves the empty $\ket{0}$ and singly occupied states $\ket{\uparrow}$ and $\ket{\downarrow}$ unchanged.  For time-reversal, $u_{\mathcal{T}} = \text{exp} \left( i \pi S^y \right)$ and $\Gamma_{j'}$ is replaced by $\Gamma_{j'}^*$ (complex conjugate). Finally, for the dihedral transformation, $u_{\mathcal{D}_2} = \text{exp} \left( i \pi S^x \right) \times  \text{exp} \left( i \pi S^z \right)$ and $\Gamma_{j'}$ remains the same.

Eq. (\ref{eqn:mps-transform}) implies that the Schmidt eigenstates of the left and right halves of the system transforms under symmetry operation $u_g$ as
\begin{eqnarray}
u_g \ket{\alpha}_L = \sum_\beta \left( U_g \right)_{\beta \alpha} \ket{\beta}_L
\label{eqn:mps-transform2}
\end{eqnarray}
for the left part and by the conjugate matrix for the right part. This means that the Schmidt eigenstates transform according to a projective representation of the symmetry group of the system. The phases of $U_g$ are not uniquely determined by Eqs. (\ref{eqn:mps-transform}) and (\ref{eqn:mps-transform2}) and it is this phase ambiguity that determines the degeneracy of the ES. For example in the case of $\mathcal{I}$, the transformation law is given by
\begin{eqnarray}
\Gamma_j^T = e^{i \theta_\mathcal{I}} U_\mathcal{I}^\dagger \Gamma_j U_\mathcal{I}.
\label{eqn:mps-transform3}
\end{eqnarray}
Relating Eq. (\ref{eqn:mps-transform3}) to the transfer matrix Eq. (\ref{eqn:imps2}) gives
\begin{eqnarray}
\sum_j \Gamma_j^\dagger \Lambda U_\mathcal{I} U_\mathcal{I}^* \Lambda \Gamma = e^{2 i \theta_\mathcal{I}} U_\mathcal{I} U_\mathcal{I}^*,
\label{eqn:mps-transform4}
\end{eqnarray}
i.e. $U_\mathcal{I} U_\mathcal{I}^*$ is an eigenvector of $T$ with eigenvalue $e^{2i \theta_\mathcal{I}}$. Since the eigenvalue of the left and right eigenvectors of Eq. (\ref{eqn:transfer-matrix}) are set to 1, the eigenvalue of  $U_\mathcal{I} U_\mathcal{I}^*$ is also 1 and is unique. Thus by comparison, $e^{2i \theta_\mathcal{I}} = 1$ and
\begin{eqnarray}
U_\mathcal{I} U_\mathcal{I}^* = e^{i \phi_\mathcal{I}}
\label{en:sym_algebra}
\end{eqnarray}
where $\phi_\mathcal{I}$ is a phase. Iterating the latter equation twice gives $e^{2i \phi_\mathcal{I}} = 1$, i.e. $\phi_\mathcal{I} = 0$ or $\pi$.

In the SPT phase, $\phi_\mathcal{I} = \pi$, thus $U_\mathcal{I}$ is an antisymmetric matrix and the eigenvalues $\Lambda_\alpha$ are at least 2-fold degenerate. More generally, since $U_\mathcal{I}$ transforms the $k_\alpha$-dimensional subspace of states with eigenvalue $\Lambda_\alpha$ within itself, $U_\mathcal{I}^\alpha$ satisfies $\text{det} \left[ U_\mathcal{I}^\alpha \right] = \text{det} \left[ \left( U_\mathcal{I}^\alpha \right)^T \right] = \text{det}\left[ - U_\mathcal{I}^\alpha \right] = \left( -1 \right)^{k_\alpha} \text{det} \left[ U_\mathcal{I}^\alpha \right]$. Since $U_\mathcal{I}^\alpha$ is unitary, $\text{det} \left[ U_\mathcal{I}^\alpha \right] \neq 0$ and therefore $\left( -1 \right)^{k_\alpha} = 1$, i.e. the multiplicity $k_\alpha$ is constrained to even integers. In the topologically trivial phase, $\phi_\mathcal{I} = 0$ and $U_\mathcal{I}$ is a symmetric matrix. Thus, there is no contraint on the degeneracy of the ES. The discrete nature of the values that $\phi_\mathcal{I}$ can take indicates that $\phi_\mathcal{I}$ cannot change unless a phase transition occurs. This is because at the critical point, the transfer matrix $T$ contains a pair of unimodular eigenvectors and this causes the correlation length to diverge, hence $U_\mathcal{I} U_\mathcal{I}^*$ is not defined.

An example of $\mathcal{I}$ protecting the SPT phase can be shown for the AKLT state which is a state in the Haldane phase \cite{Pollman}. Writing the AKLT state in an MPS form with $\Gamma_a = \sqrt{\frac{2}{3}} \sigma_a$ and $\Lambda = \frac{1}{\sqrt{2}} \mathbbm{1}$, where  $\sigma_a (a = x,y,z)$ are the Pauli matrices, it can be shown that under inversion, $\sigma_a \rightarrow \sigma_a^T = - \sigma_y \sigma_a \sigma_y$ and one obtains $U_\mathcal{I} = \sigma_y$ and $\theta_\mathcal{I} = \pi$. Also since $U_\mathcal{I} U_\mathcal{I}^* = \sigma_y \sigma_y^* = -\mathbbm{1}$, one finds $e^{i \phi_\mathcal{I}} = -1$ and $\phi_\mathcal{I} = \pi$. Thus the AKLT state can be characterized by $\theta_\mathcal{I} = \pi$, $\phi_\mathcal{I} = \pi$ and the even-fold degenerate ES.

%%%%%%%%%%%%%%%%%%%%%%%%%%%%%%%%%%%%%%%%%%%%%%%%%%%%%%%%%%%%%%%%%%%%%%%%%%%%%%%%%%%%%%%%%%%%%%%%%%%%%%%%%%%
%%%%%%%%%%%%%%%%%%%%%%%%%%%%%%%%%%%%%%%%%%%%%%%%%%%%%%%%%%%%%%%%%%%%%%%%%%%%%%%%%%%%%%%%%%%%%%%%%%%%%%%%%%%

\section{Results}
\label{results}
\begin{figure}[h!]
  \centering\includegraphics[width=0.45\textwidth]{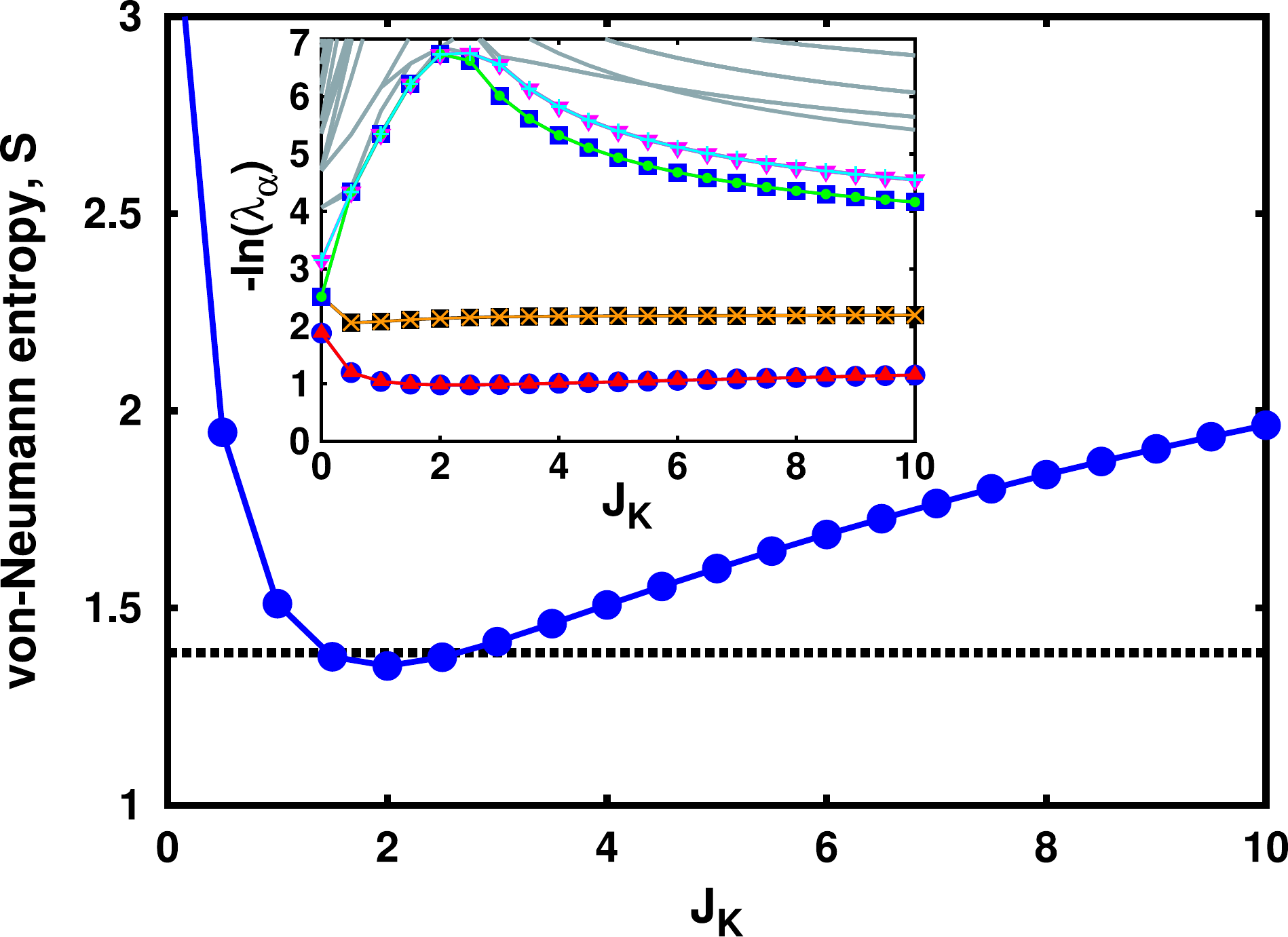}
  \caption{(Colour online) von-Neumann entropy $S$ versus \textit{p}-wave Kondo coupling strength $J_K$ with $J_\perp = U = 0$. The minimum of $S$ occurs at $J_K = 2$ and increases away from this due fermion hopping and local spin fluctuations when $J_K < 2$, and non-local fermion hopping when $J_K > 2$. The minimum of $S$ occurs below $2 \ln 2$ (horizontal dashed line) indicating that $S$ is directly caused by the 4-fold degeneracy of 2 free edge spin-1/2's as assumed in Ref. \cite{Lisandrini}. Inset: Low-lying ES versus $J_K$. The different coloured symbols represent different low-lying ES values while the grey lines are the higher ES values. The entire spectrum in even-fold degenerate, indicating an SPT phase.}
  \label{fig:svn-J_K}
\end{figure}

Throughout this work, the parameters $t$ and $J_H$ are set to unity and the Hubbard chain is half-filled. All data presented are for a translationally invariant unit cell of two sites - one itinerant fermion site on the Hubbard chain and one local spin-1/2 site on the Heisenberg chain. When $U = J_\perp = 0$ and $J_K > 0$, the groundstate of the TKI is known to be in the Haldane phase \cite{Mezio}. The Hamiltonian and groundstate wavefunction in this phase contain the symmetries $\mathcal{I}$, $\mathcal{T}$ and $\mathcal{D}_2$. The Haldane phase however is only protected by $\mathcal{I}$ since it is the only symmetry of the three that is not graded. This is confirmed by the ``non-local" order parameter $O_\mathcal{I} = \braket{U_\mathcal{I} U_\mathcal{I}^*} = -1$ throughout the entire range of $J_K$. This gives the phase $\phi_\mathcal{I} = \pi$ which remains constant throughout the range of $J_K$ which contributes to an even-fold degeneracy of the ES shown in the inset of Fig. \ref{fig:svn-J_K}. The different colours and symbols in the inset of Fig. \ref{fig:svn-J_K} indicate the different low-lying ES values while the grey lines are the higher ES values. Since the representations of $\mathcal{T}$ and $\mathcal{D}_2$ are graded, $|O_\mathcal{T}| < 1$ and $|O_{\mathcal{D}_2}| < 1$.

Fig. \ref{fig:svn-J_K} shows the von-Neumann entropy versus $J_K$. The minimum of $S$ occurs at $J_K \approx 2$ with a value of $S \approx 2 \ln 2$. In Ref. \cite{Lisandrini}, this value of $S$ was assumed to be caused by the 4-fold groundstate degeneracy due to the 2 free edge spin-1/2's and served as an indicator of the Haldane phase. This assumption is disproved by the work done here where it will be shown in Section \ref{Coulomb-svn1} that it is possible to lower this minimum of $S$ while still being in the Haldane phase. For $J_K < 2$, the Hubbard and Heisenberg chains are weakly coupled, leading to a large degree of freedom in the two chains, and thus a diverging $S$. For $J_K > 2$, $H_K$ dominates the other Hamiltonian terms and $S$ increases. This could be understood from the non-local fermion hopping term contained in $H_K$ by expressing it as $H_K = H_1 + H_2$ where
\begin{eqnarray}
H_1 &=& \frac{J_K}{2} \sum_j \left[ \frac{1}{2} \left\lbrace S^+_j \left( s^-_{j+1} + s^-_{j-1} \right) + S^-_j \left( s^+_{j+1} + s^+_{j-1} \right) \right\rbrace \nonumber \right.
\\
&& \left. + S^z_j \left( s^z_{j+1} + s^z_{j-1} \right) \right]
\label{eqn_Hamiltonian_kondo_H1}
\end{eqnarray}
describes the spin exchange between a local spin-1/2 at site $j$ and the spin degree of freedom of fermions at sites $j-1$ and $j+1$. The lower case operators $s_j$ act on the spin degree of freedom of the fermions and are given by
\begin{eqnarray}
s^+_j = c^\dagger_{j, \uparrow} c_{j, \downarrow} \nonumber \quad , \quad s^-_j = c^\dagger_{j, \downarrow} c_{j, \uparrow} \nonumber ,
\\
%s^z_j = c^\dagger_{j, \uparrow} c_{j, \uparrow} \nonumber \quad , \quad -s^z_j = c^\dagger_{j, \downarrow} c_{j, \downarrow} \nonumber.
s^z_j = \frac{1}{2} \left( c^\dagger_{j, \uparrow} c_{j, \uparrow} - c^\dagger_{j, \downarrow} c_{j, \downarrow} \right) \nonumber.
\label{eqn:s_cc_ops}
\end{eqnarray}
The second term
\begin{eqnarray}
H_2 &=& -\frac{J_K}{4} \sum_j \left[ S^+_j \left( c^\dagger_{j+1, \downarrow} c_{j-1, \uparrow} + c^\dagger_{j-1, \downarrow} c_{j+1, \uparrow} \right) \nonumber \right.
\\
&& \left. + S^-_j \left( c^\dagger_{j+1, \uparrow} c_{j-1, \downarrow} + c^\dagger_{j-1, \uparrow} c_{j+1, \downarrow} \right) \nonumber \right. 
\\
&& \left. + S^z_j \left\lbrace \left( c^\dagger_{j+1, \uparrow} c_{j-1, \uparrow} + c^\dagger_{j-1, \uparrow} c_{j+1, \uparrow} \right) \nonumber \right. \right.
\\
&& \left. \left. - \left( c^\dagger_{j+1, \downarrow} c_{j-1, \downarrow} + c^\dagger_{j-1, \downarrow} c_{j+1, \downarrow} \right) \right\rbrace \right] ,
\label{eqn_Hamiltonian_kondo_H2}
\end{eqnarray}
describes the interaction between a local spin-1/2 at site $j$ and a fermion hopping from site $j \pm 1$ to site $j \mp 1$, accompanied by a spin flip. This non-local hopping term adds to the fermion hopping when $J_K$ is large and thus tends to increase the entropy.

\begin{figure}[h!]
  \centering\includegraphics[width=0.45\textwidth]{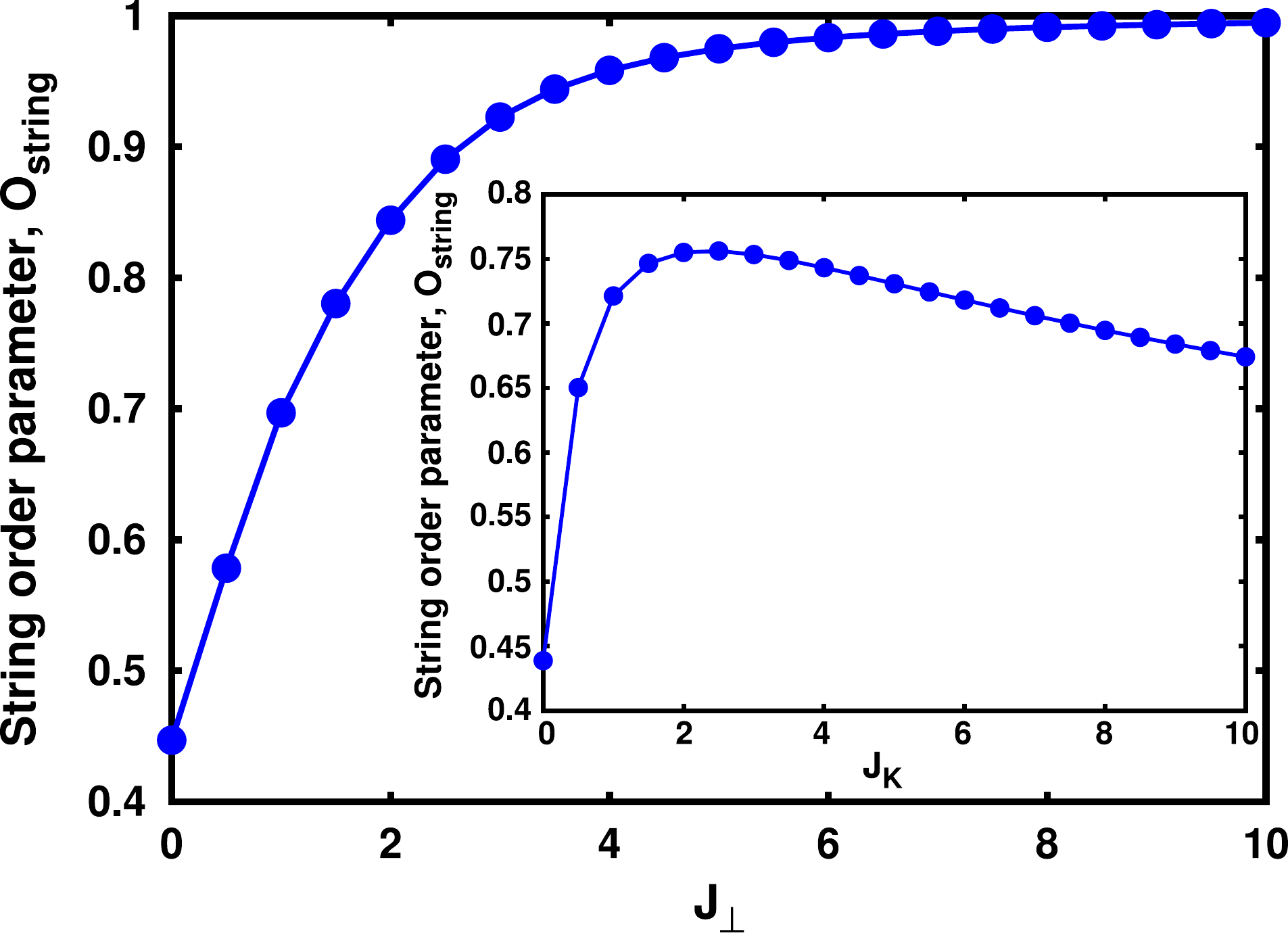}
  \caption{String order parameter $O_\text{string}$ versus $J_\perp$ with $J_K = U = 0$. The rapid increase of $O_\text{string}$ is caused by the formation of local singlets between the fermions and local spin-1/2's which reduces fermion hopping and forces fermions to take a one-charge per site order. Inset: $O_\text{string}$ versus $J_K$ with $J_\perp = U = 0$. $O_\text{string}$ peaks at $J_K = 2$ where fermions are maximally ordered with a one-charge per site occupation. When $J_K < 2$ and $J_K > 2$, fermion hopping is large and this reduces the one-charge per site order, causing $O_\text{string}$ to decrease.}
  \label{fig:ord-string2}
\end{figure}

Fig. \ref{fig:ord-string2} shows the string order parameter defined as
\begin{eqnarray}
O_\text{string}^2 \equiv \lim_{|j-k| \rightarrow \infty} \left\langle  \mathbbm{1}_j \text{exp} \left[ \frac{i \pi}{2} \sum^k_{l=j} \left( \hat{n}_l - 1 \right) \right] \mathbbm{1}_k \right\rangle ,
\label{eqn:string_order}
\end{eqnarray}
where $\hat{n}_l = \sum_\sigma c^\dagger_{l, \sigma} c_{l, \sigma}$. Due to the factor $\frac{1}{2}$ in the exponent, this string order parameter detects a 2-particle fluctuation with total spin zero in the region between sites $j$ and $k$, which is in contrast to the conventional 1-particle fluctuation with total spin half where the factor in the exponent would be unity \cite{Nijs,Anfuso,Montorsi}. This spinless 2-particle fluctuation is chosen over the 1-particle fluctuation, i.e. the formation of a fermion, in order to detect the phase transition that occurs when the charge gap vanishes while the spin gap remains finite (see Section \ref{Coulomb-svn2}). An insulating phase would have one charge per site on average, i.e. a large distribution of $\hat{n}_l = 1$ as compared to $\hat{n}_l = 0$ and 2. Thus, $\sum^k_{l=j} \left( \hat{n}_l - 1 \right) \rightarrow 0$ as $|j-k| \rightarrow \infty$. This causes $\text{exp} \left[ \frac{i \pi}{2} \sum^k_{l=j} \left( \hat{n}_l - 1 \right) \right] \rightarrow 1$ and $O_\text{string} \rightarrow 1$. In the opposite case where $\hat{n}_l = 0$ and 2 \emph{originating from a two particle fluctuation} outweighs $\hat{n}_l = 1$, the sum $\sum^k_{l=j} \left( \hat{n}_l - 1 \right) \rightarrow -2m$ for $\hat{n}_l = 0$ and $\sum^k_{l=j} \left( \hat{n}_l - 1 \right) \rightarrow 2m$ for $\hat{n}_l = 2$, where $m$ is the number of \emph{pairs} of empty or doubly-occupied sites. The exponents however give the same value $\text{exp} \left[ \frac{i \pi}{2} \sum^k_{l=j} \left( \hat{n}_l - 1 \right) \right] \rightarrow -1$ for both $\hat{n}_l = 0$ and 2, hence $O_\text{string} \rightarrow -1$. Details of the evaluation of $O_\text{string}$ in an iMPS is shown in Section \ref{appendix_Ostring}.

The inset of Fig. \ref{fig:ord-string2} shows $O_\text{string}$ versus $J_K$ with $J_\perp = U = 0$. $O_\text{string}$ peaks at $J_K \approx 2 - 2.5$ where $O_\text{string} \approx 0.75$, indicating that most of the charges in the groundstate have a one particle per site order. When $J_K < 2$, $O_\text{string}$ decreases due to fermion hopping originating from the Hubbard term in Eq. (\ref{eqn_Hamiltonian_Hubbard}), while when $J_K > 2.5$, $O_\text{string}$ decreases but this time the fermion hopping comes from the non-local hopping term Eq. (\ref{eqn_Hamiltonian_kondo_H2}). The occurrence of both the minimum of $S$ and the maximum of $O_\text{string}$ at $J_K \approx 2$ shows that the groundstate is in it's maximum ordered, insulating phase, which separates two more disordered, weaker insulating (more metallic) phases.

\begin{figure}[h!]
  \centering\includegraphics[width=0.45\textwidth]{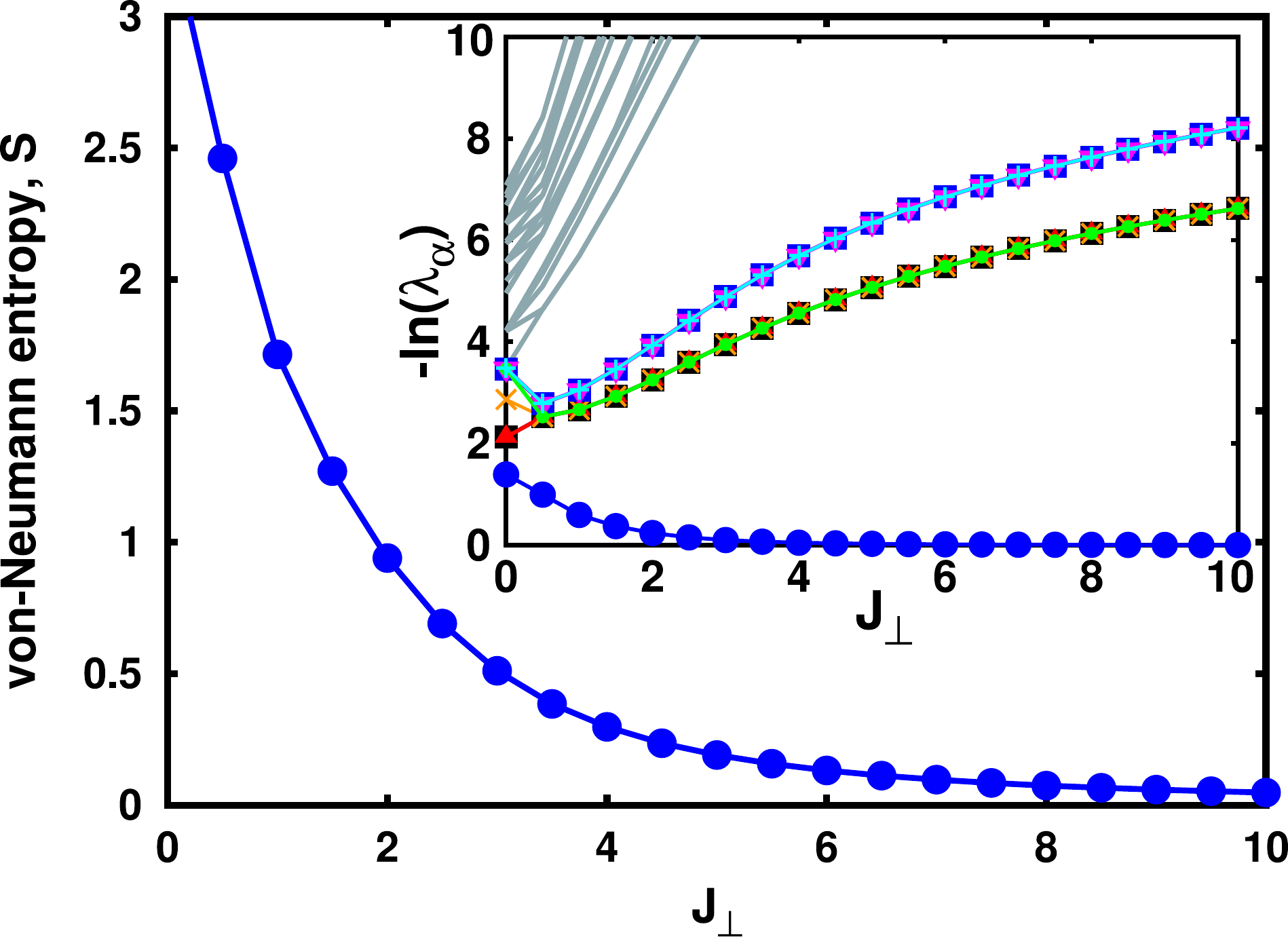}
  \caption{(Colour online) von-Neumann entropy $S$ versus \textit{s}-wave Kondo coupling strength $J_\perp$ with $J_K = U = 0$. $S$ is not lower-bounded and decreases as $J_\perp$ increases due to the formation of local singlets between fermions and local spin-1/2's. The groundstate is a direct product of local singlets which has low entanglement. Inset: Low-lying entanglement spectrum as a function of $J_\perp$. The different coloured symbols represent different low-lying ES values while the grey lines are the higher ES values. The lowest ES value (blue circle) is non-degenerate, indicating that the groundstate is topologically trivial.}
  \label{fig:svn-Jperp}
\end{figure}

In the opposite case where $J_K = U = 0$ and $J_\perp > 0$, all three symmetries $\mathcal{I}$, $\mathcal{T}$ and $\mathcal{D}_2$ are still present in the Hamiltonian and groundstate wavefunction. However, $O_\mathcal{I} = \braket{U_\mathcal{I} U_\mathcal{I}^*} = 1$ and this gives $\phi_\mathcal{I} = 0$ throughout the range of $J_\perp$. This causes the ES (inset of Fig. \ref{fig:svn-Jperp}) of the groundstate wavefunction to contain both even- and odd-fold degenerate eigenvalues which implies that the groundstate is topologically trivial. Since the representations of $\mathcal{T}$ and $\mathcal{D}_2$ are graded, $|O_\mathcal{T}| < 1$ and $|O_{\mathcal{D}_2}| < 1$.

Just as in the previous case, when $J_\perp \rightarrow 0$, the Hubbard and Heisenberg chains are decoupled from one another, thus there is a large degree of freedom due to the fluctuations of the free fermions and local spin-1/2's. This causes the von-Neumann entropy $S$ in Fig. \ref{fig:svn-Jperp} to diverge and $O_\text{string}$ in the main plot of Fig. \ref{fig:ord-string2} to decrease. When $J_\perp \neq 0$, the local Kondo interaction $H_\perp$ quickly overcomes the fermion hopping and local spin-1/2 fluctuations by forming local singlets. This causes $O_\text{string}$ to increase rapidly and $S$ to decay exponentially. As $J_\perp \rightarrow \infty$, one would expect $O_\text{string} \rightarrow 1$ and $S \rightarrow 0$ since the groundstate is comprised of a direct product of local singlets, i.e. zero fermion hopping and local spin fluctuations.

%%%%%%%%%%%%%%%%%%%%%%%%%%%%%%%%%%%%%%%%%%%%%%%%%%%%%%%%%%%%%%%%%%%%%%%%%%%%%%%%%%%%%%%%%%%%%%%%%%%%%%%%%%%
%%%%%%%%%%%%%%%%%%%%%%%%%%%%%%%%%%%%%%%%%%%%%%%%%%%%%%%%%%%%%%%%%%%%%%%%%%%%%%%%%%%%%%%%%%%%%%%%%%%%%%%%%%%

\subsection{Effect of Hubbard interaction, $U \neq 0$}
\label{Coulomb-svn1}
\begin{figure}[h!]
  \centering\includegraphics[width=0.45\textwidth]{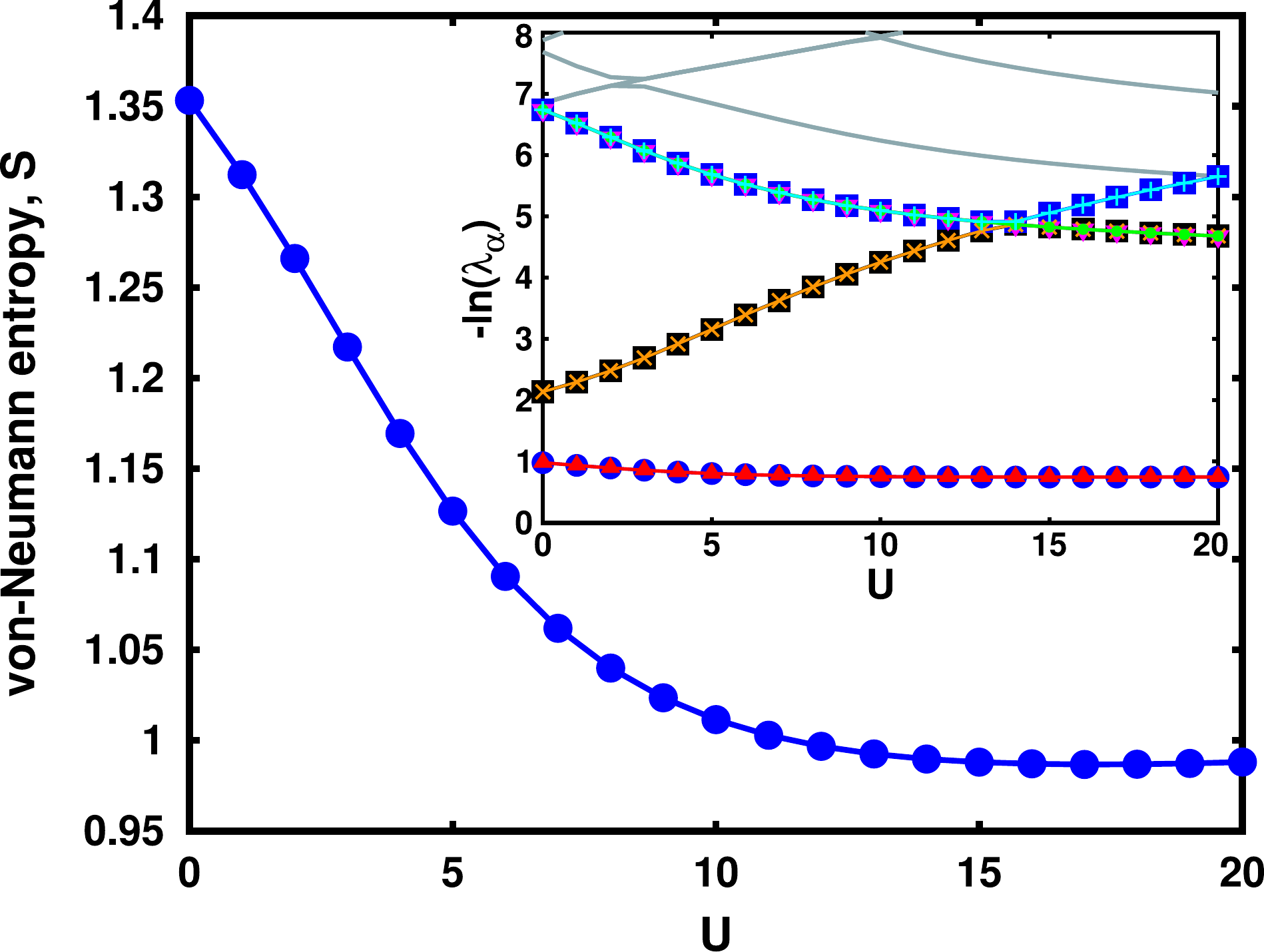}
  \caption{(Colour online) von-Neumann entropy $S$ versus Hubbard interaction $U$ with $J_K = 2$ and $J_\perp = 0$. Increasing $U$ reduces fermion hopping and this causes $S$ to decrease. When $U > 10$, $S$ evens out but does not vanish due to the spin degree of freedom of the fermions originating from the non-local $p$-wave coupling. Inset: Low-lying ES versus $U$. The different coloured symbols represent different low-lying ES values while the grey lines are the higher ES values. The entire ES is even-fold degenerate, indicating that the groundstate is still in an SPT phase.}
  \label{fig:svn-U-J_K}
\end{figure}

The effect of the Hubbard interaction in Eq. (\ref{eqn_Hamiltonian_Hubbard}) is to energetically penalize the system when there is more than 1 fermion with opposite spins per site. This means in order to lower its energy, the system will prefer a configuration where there is only one fermion per site, i.e. fermion hopping gets suppressed and charge fluctuations are frozen out when $U \rightarrow \infty$.

\begin{figure}[h!]
  \centering\includegraphics[width=0.45\textwidth]{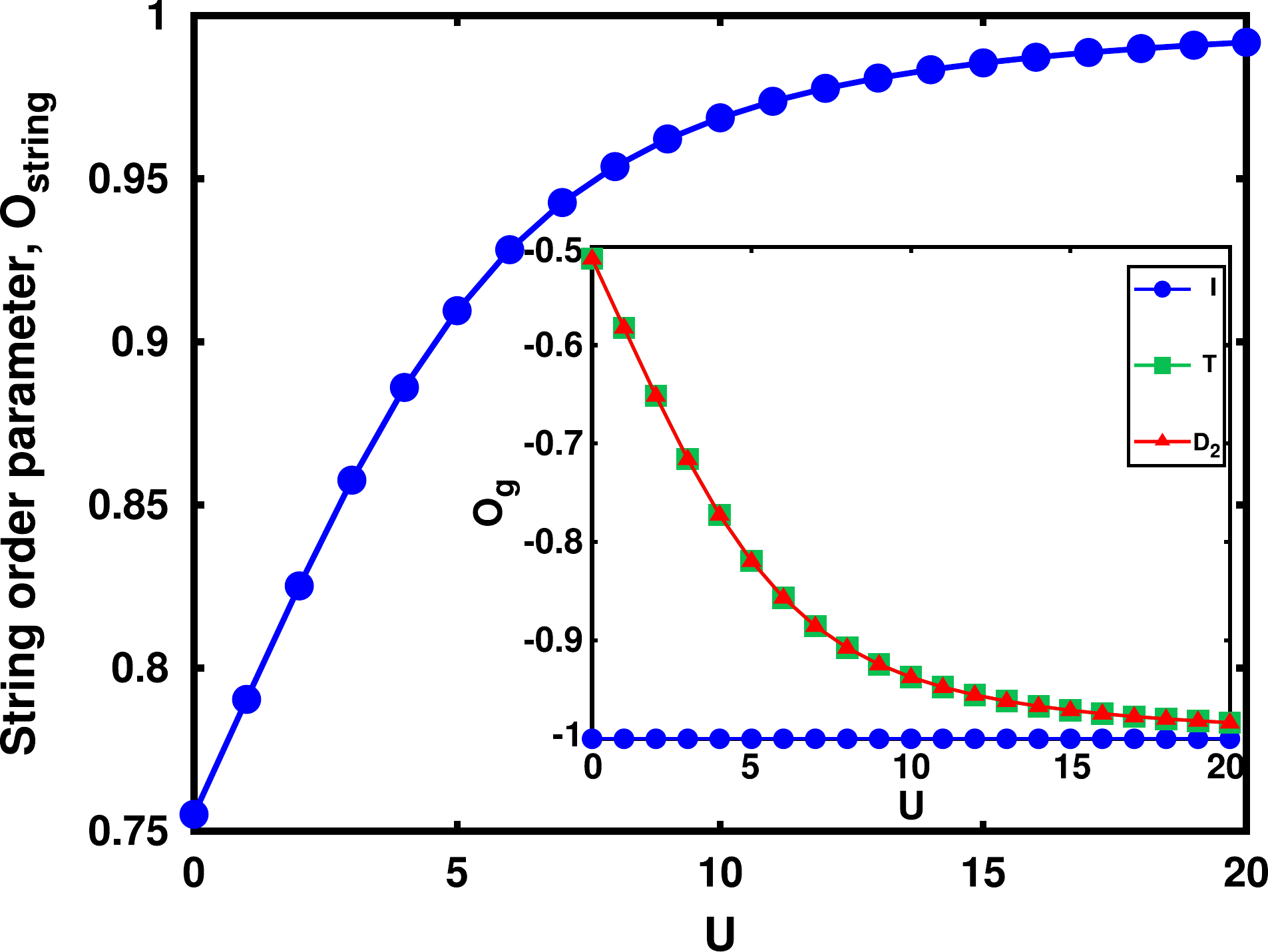}
  \caption{(Colour online) $O_\text{string}$ versus Hubbard interaction $U$ with $J_K = 2$ and $J_\perp = 0$. As $U$ increases, fermions repel more strongly and are forced to take a one-charge per site occupation, causing $O_\text{string}$ to increase. Inset: Non-local order parameters $O_g$ versus $U$ where $g = \mathcal{I}$ (blue circles), $\mathcal{T}$ (green squares) and $\mathcal{D}_2$ (red triangles). $O_\mathcal{I} = -1$ indicates that the Haldane phase is protected by $\mathcal{I}$, whereas $|O_\mathcal{T}| < 1$ and $|O_{\mathcal{D}_2}| < 1$ indicate that $\mathcal{T}$ and $\mathcal{D}_2$ are graded and therefore are not protecting symmetries of the Haldane phase.}
  \label{fig:ord-U-J_K}
\end{figure}

Fig. \ref{fig:svn-U-J_K} shows the von-Neumann entropy $S$ versus Hubbard interaction $U$ with $J_K = 2$ and $J_\perp = 0$. $S$ decreases with increasing $U$ and tends to a non-zero constant for large $U$, indicating that $S$ is lower-bounded (See Section \ref{appendix_largeU} for the case of the large $U$ limit). The decrease of $S$ is caused by the reduction of fermion hopping as the Hubbard interaction forces fermions to occupy single sites. This can be seen in the increase of $O_\text{string}$ in Fig. \ref{fig:ord-U-J_K}. As $U$ is further increased, $O_\text{string}$ changes slowly and tends to 1 as $U \rightarrow \infty$ where each site contains only 1 fermion. This however does not cause $S$ to vanish completely, instead the non-zero lower-bound contribution to $S$ comes from the non-local interaction between the spin degree of freedom of the frozen fermions and the local spin-1/2's originating from the non-local $p$-wave coupling in $H_K$. This entanglement contribution is not affected by further increasing $U$ since the Hubbard interaction does not affect the spin degree of freedom of the frozen fermions.

The inset of Fig. \ref{fig:ord-U-J_K} shows the non-local order parameters defined in Eq. (\ref{eqn:non-local-ord-param}). As explained earlier, $\mathcal{I}$ is responsible for protecting the topological phase, thus $O_\mathcal{I} = -1$ throughout the entire range of $U$ since the Hubbard interaction has no effect on the spatial inversion of the system. This causes the even-fold degeneracy of the ES values shown in the inset of Fig. \ref{fig:svn-U-J_K}. In contrast to $O_\mathcal{I}$, the other two non-local string order parameters $O_\mathcal{T}$ and $O_{\mathcal{D}_2}$ decrease with increasing $U$. Since these two quantities are still decreasing at $U = 20$, one can expect that they both tend to -1 when $U \rightarrow \infty$. In this limit, all charge fluctuations are frozen out and the Hubbard chain is effectively equal to a Heisenberg chain consisting of local spin-1/2's. In such a case, $\mathcal{T}$ and $\mathcal{D}_2$ are no longer graded and they become protecting symmetries of the Haldane phase.

\begin{figure}[h!]
  \centering\includegraphics[width=0.45\textwidth]{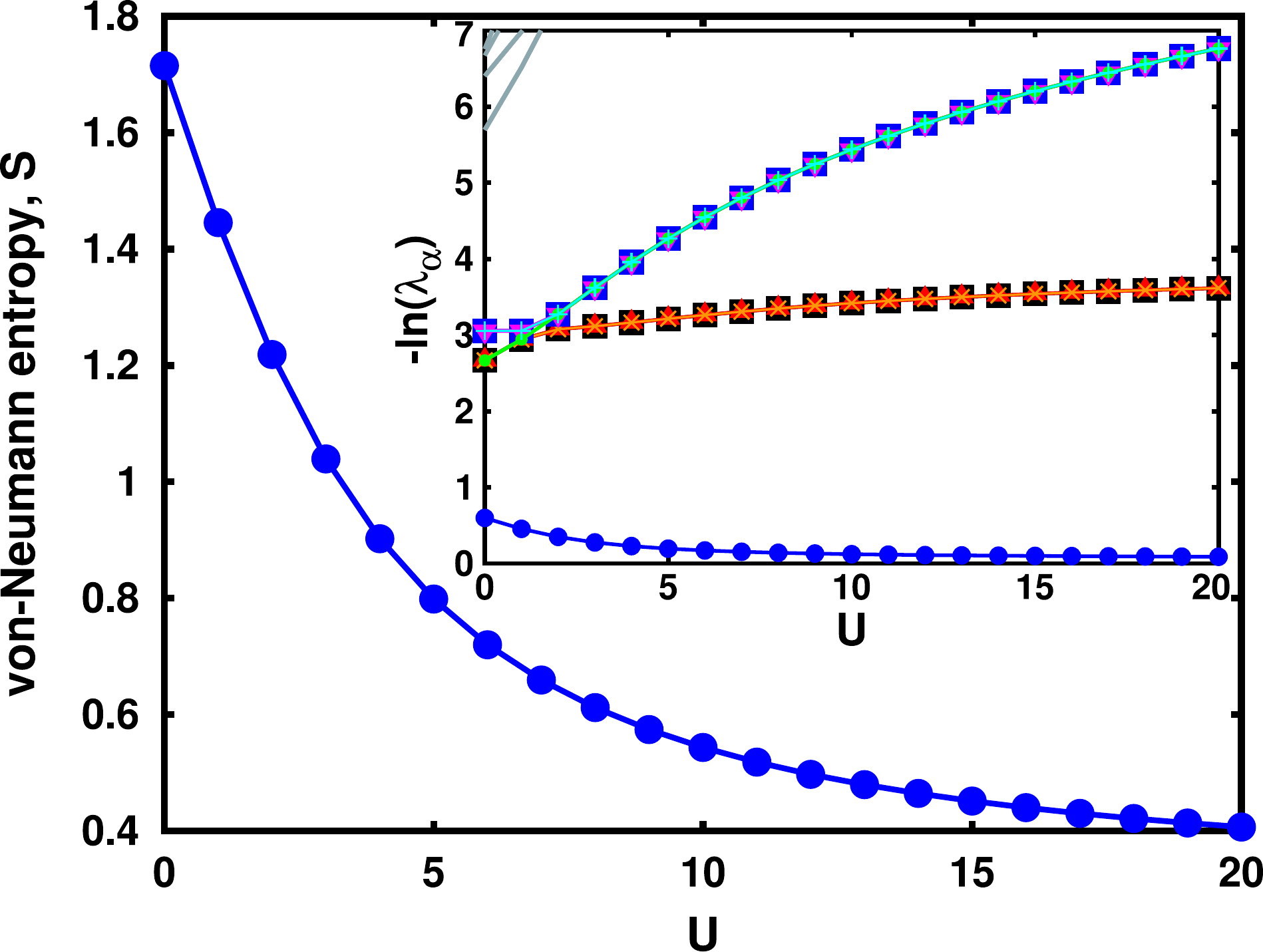}
  \caption{(Colour online) von-Neumann entropy $S$ versus Hubbard interaction $U$ with $J_K = 0$ and $J_\perp = 1$. Increasing $U$ reduces fermion hopping and forces fermions to take a one-charge per site occupancy, hence decreasing $S$. Unlike Fig. \ref{fig:svn-U-J_K}, $S$ is not lower bounded since there is no non-local fermion hopping, thus it is possible for $S$ to vanish when $U \rightarrow \infty$. Inset: Low-lying ES versus $U$. The different coloured symbols represent different low-lying ES values while the grey lines are the higher ES values. The lowest ES value is non-degenerate, indicating a topologically trivial phase.}
  \label{fig:svn-U-Jperp}
\end{figure}

Fig. \ref{fig:svn-U-Jperp} shows the von-Neumann entropy $S$ versus $U$ when $J_\perp = 1$ and $J_K = 0$. Since $S$ still appears to be decreasing when $U = 20$, one can expect that $S \rightarrow 0$ as $U \rightarrow \infty$ (See Section \ref{appendix_largeU} for the case of the large $U$ limit). This can be understood by analyzing the groundstate structure favored by $H_c$ in Eq. (\ref{eqn_Hamiltonian_Hubbard}) and $H_\perp$ in Eq. (\ref{eqn_Hamiltonian_Kondo_local}) when $U \gg t$. The effect of the former is to energetically penalize the system when more than one fermion occupies a site while the latter binds a fermion to a local spin-1/2 at site $j$ to form local singlets. Ultimately, both these effects favor a groundstate that contains one fermion per site. This type of groundstate consists of trivial products of local pairs consisting of a fermion and a local spin-1/2, which has low or no entanglement with neighbouring pairs. As a result, there is no lower bound to $S$ and $S \rightarrow 0$ as $U \rightarrow \infty$.

\begin{figure}[h!]
  \centering\includegraphics[width=0.45\textwidth]{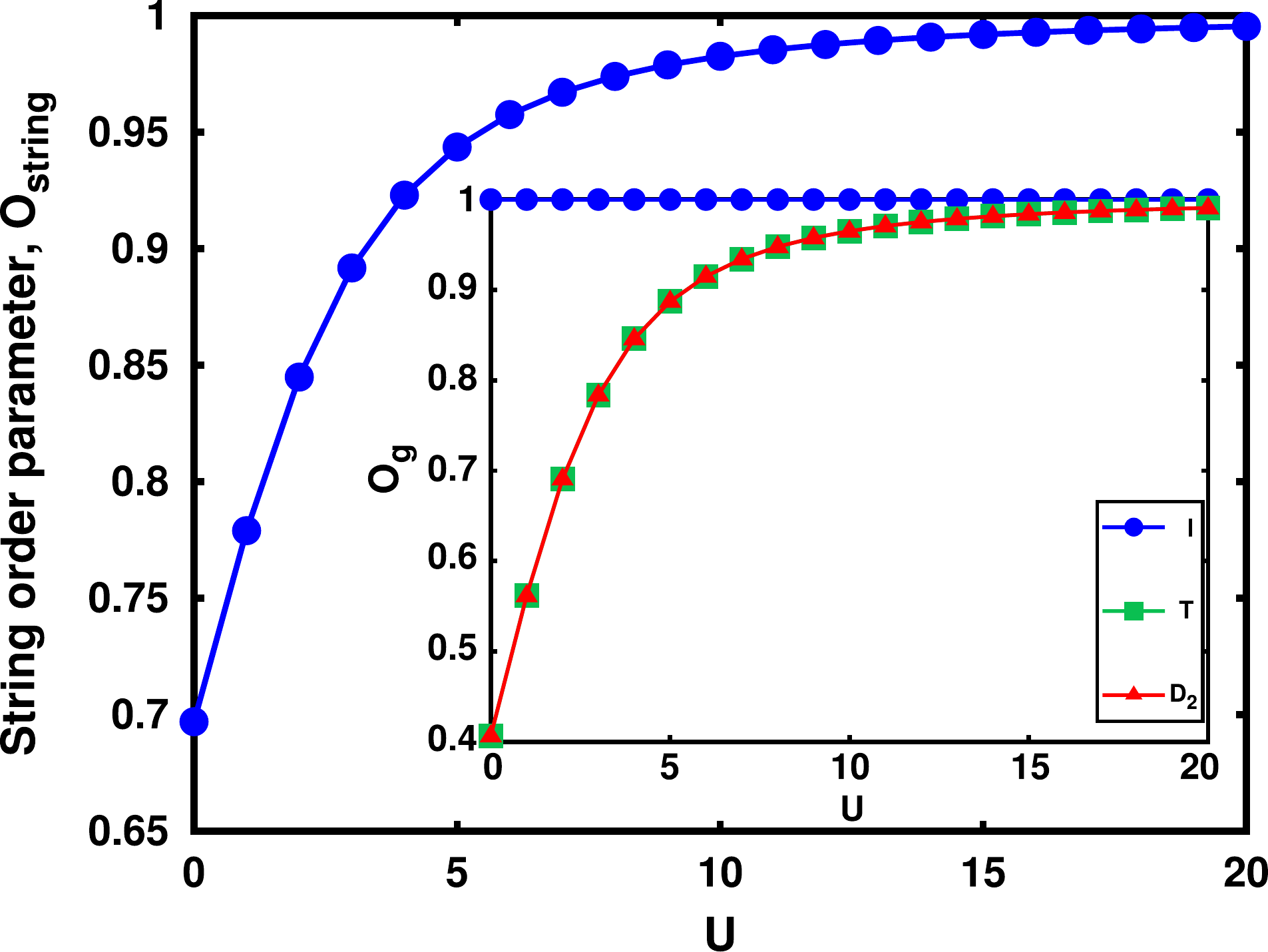}
  \caption{(Colour online) $O_\text{string}$ versus Hubbard interaction $U$ with $J_K = 0$ and $J_\perp = 1$. As $U$ increases, fermions repel more strongly and are forced to take a one-charge per site occupation, causing $O_\text{string}$ to increase. Inset: Non-local order parameters $O_g$ versus $U$ where $g = \mathcal{I}$ (blue circles), $\mathcal{T}$ (green squares) and $\mathcal{D}_2$ (red triangles). $O_\mathcal{I} = 1$ indicates that the groundstate is topologically trivial and is symmetric under inversion. $|O_\mathcal{T}| < 1$ and $|O_{\mathcal{D}_2}| < 1$ indicate that $\mathcal{T}$ and $\mathcal{D}_2$ are graded.}
  \label{fig:ord-U-Jperp}
\end{figure}

Since $J_K = 0$, there is no additional fermion hopping originating from the non-local $p$-wave coupling, therefore the effect of $U$ on $O_\text{string}$ is greater than the case with $J_K \neq 0$ as can be seen in Fig. \ref{fig:ord-U-Jperp}. The inset of Fig. \ref{fig:ord-U-Jperp} shows the non-local order parameters as a function of $U$. $\mathcal{I}$ is not affected by $U$ and since the system is topologically trivial, $O_\mathcal{I} = 1$. This causes the non-degeneracy of the lowest ES in the inset of Fig. \ref{fig:svn-U-Jperp}. Similar to the previous case of $J_K = 2$ and $J_\perp = 0$, $\mathcal{T}$ and $\mathcal{D}_2$ are graded. As $U \rightarrow \infty$, $O_\mathcal{T}$ and $O_{\mathcal{D}_2}$ tend to 1 because charge fluctuations are frozen out and these two symmetries are no longer graded.

%%%%%%%%%%%%%%%%%%%%%%%%%%%%%%%%%%%%%%%%%%%%%%%%%%%%%%%%%%%%%%%%%%%%%%%%%%%%%%%%%%%%%%%%%%%%%%%%%%%%%%%%%%%
%%%%%%%%%%%%%%%%%%%%%%%%%%%%%%%%%%%%%%%%%%%%%%%%%%%%%%%%%%%%%%%%%%%%%%%%%%%%%%%%%%%%%%%%%%%%%%%%%%%%%%%%%%%

\subsection{Topological phase transition with $U = 0$}
\label{Coulomb-svn2}
\begin{figure}[h!]
  \centering\includegraphics[width=0.45\textwidth]{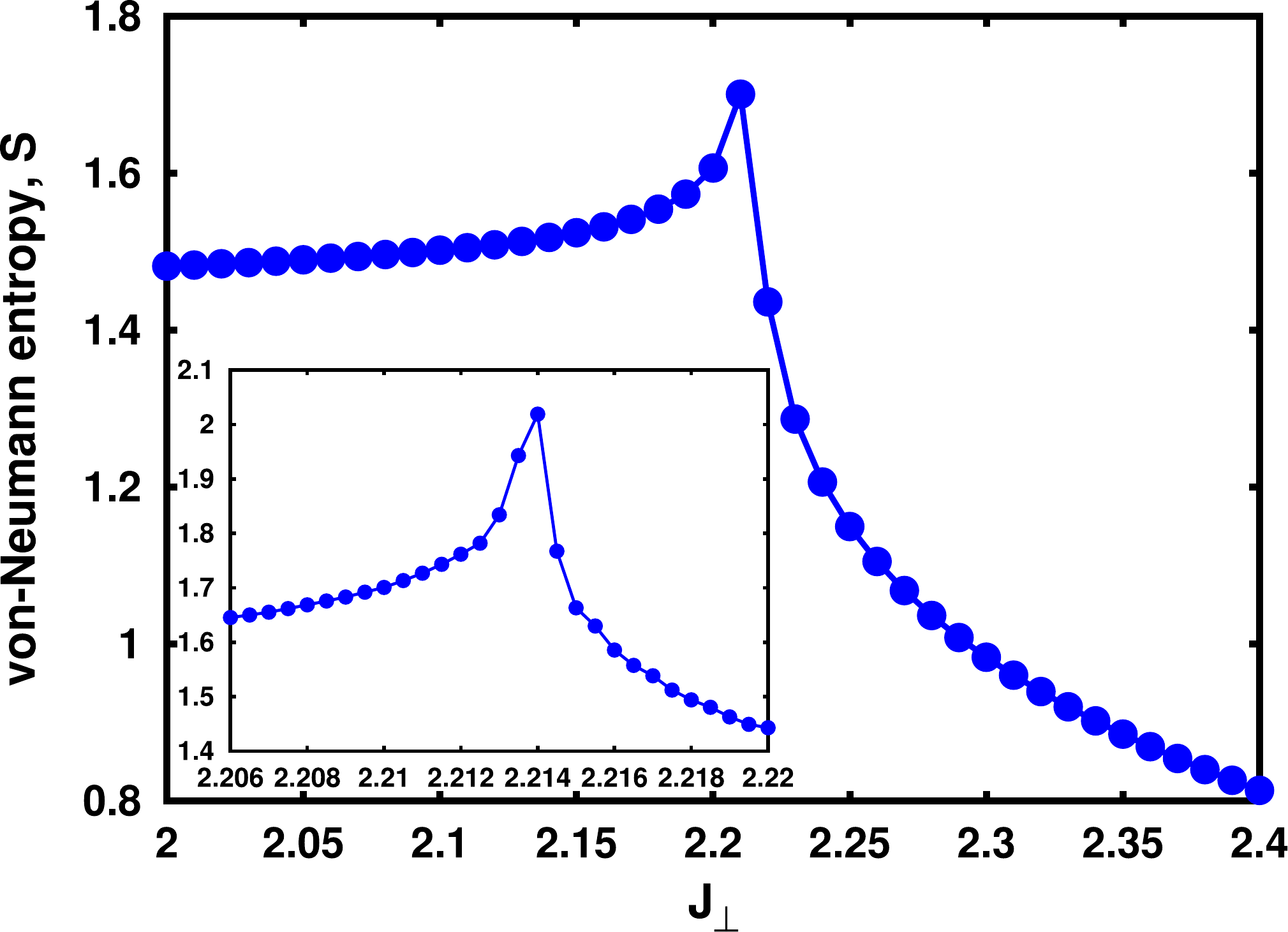}
  \caption{von-Neumann entropy $S$ versus $J_\perp$ with $J_K = 2$ and $U = 0$. $S$ diverges at the critical point $J_\perp^c$ where the topological phase transition occurs. The groundstate is in the SPT phase when $J_\perp < J_\perp^c$ and in the topologically trivial phase when $J_\perp > J_\perp^c$. Inset: $S$ in the range $J_\perp = $ 2.206 - 2.22. The critical point obtained here is $J_\perp^c = 2.2140 \pm 5 \times 10^{-4}$}
  \label{fig:svn-U0-JK-Jperp}
\end{figure}

\begin{figure}[h!]
  \centering\includegraphics[width=0.45\textwidth]{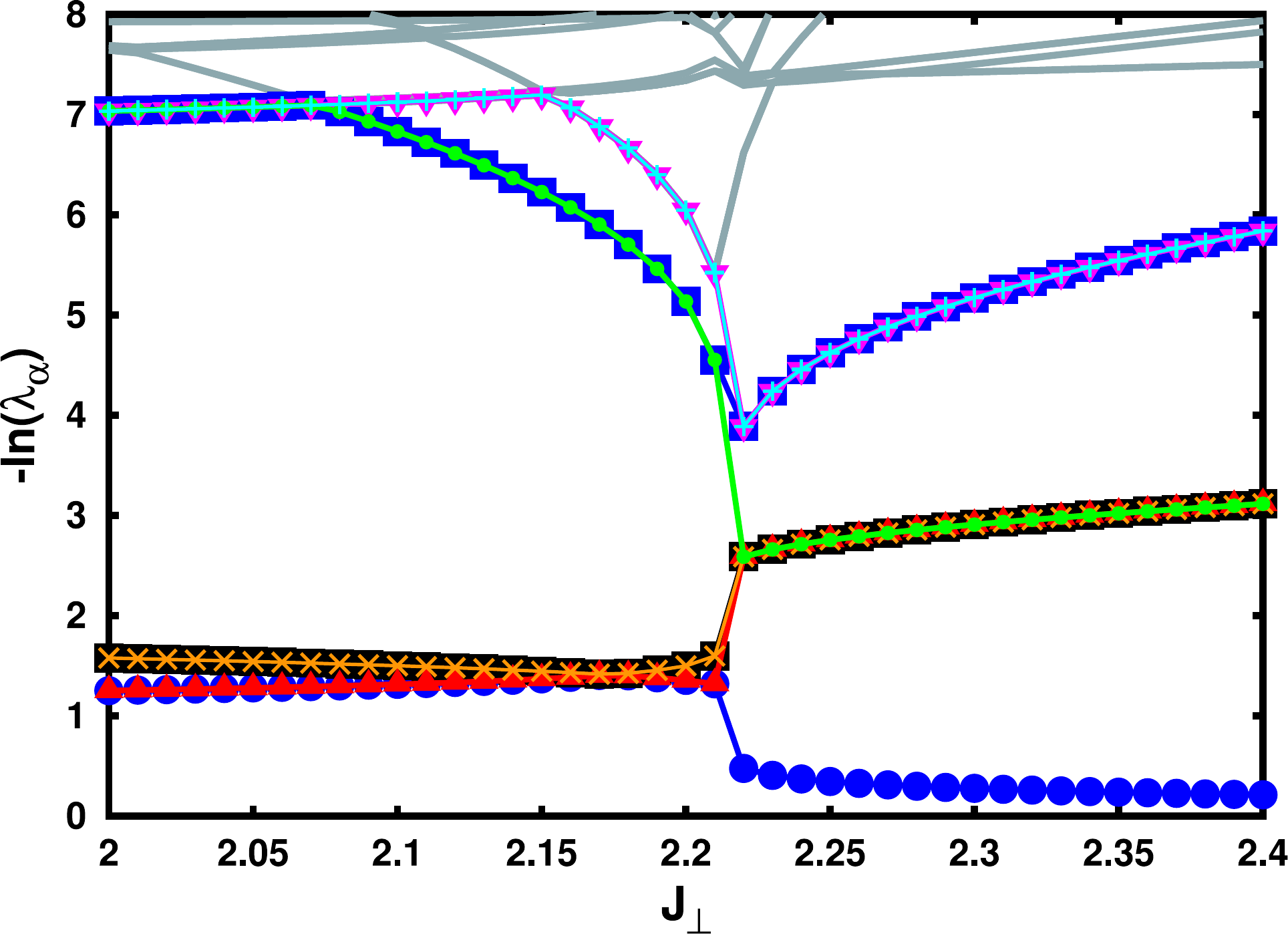}
  \caption{(Colour online) Low-lying ES values versus $J_\perp$ with $J_K = 2$ and $U = 0$. The different coloured symbols represent different low-lying ES values while the grey lines are the higher ES values. All ES values are even-fold degenerate when $J_\perp < J_\perp^c$, indicating the groundstate is in an SPT phase. When $J_\perp > J_\perp^c$, the lowest ES value is non-degenerate and the groundstate is topologically trivial. The discontinuity of the ES values occur at critical point $J_\perp^c$ where the even fold-degeneracy constraint of the ES values are lifted.}
  \label{fig:ent-spec}
\end{figure}

In this section, the two following parameters are held fixed at $J_K = 2$ and $U = 0$ while $J_\perp$ is varied. The topological phase transition occurs at the critical point $J_\perp^c = 2.2140 \pm 5 \times 10^{-4}$ where $S$ in Fig. \ref{fig:svn-U0-JK-Jperp} diverges and the lowest ES values in Fig. \ref{fig:ent-spec} changes from two-fold degenerate to non-degenerate. When $J_\perp < J_\perp^c$, $S$ is lower-bounded and the ES values shown in Fig. \ref{fig:ent-spec} are even-fold degenerate, indicating a non-trivial topological phase. When $J_\perp > J_\perp^c$, $S$ decays and has no lower bound while the lowest ES value is non-degenerate, indicating a topologically trivial phase. The inset of Fig. \ref{fig:svn-U0-JK-Jperp} shows this phase transition in the vicinity of $J^c_\perp$. At $J_\perp \approx 2.15 - 2.20$, the first lowest pair (blue circle and red triangle) and second lowest pair (black square and amber cross) of degenerate ES values in Fig. \ref{fig:ent-spec} appear to merge. However, upon close inspection, they do not merge. Instead, the two lowest pairs of degenerate ES values are simply close in value but are still distinct.

With $J_K = 2$ and $J_\perp^c = 2.2140 \pm 5 \times 10^{-4}$, the ratio of $J^c_\perp / J_K = 1.107$. A similar result was obtained in Ref. \cite{Lisandrini} where the authors used a conventional DMRG method with a fixed bond dimension of $m = 800$, and finite-size scaling of the system size to obtain $J^c_\perp / J_K = 1.11$. This difference of only $0.27\%$ is a remarkable agreement between the finite DMRG used in Ref. \cite{Lisandrini} and iDMRG used in this work.

\begin{figure}[h!]
  \centering\includegraphics[width=0.45\textwidth]{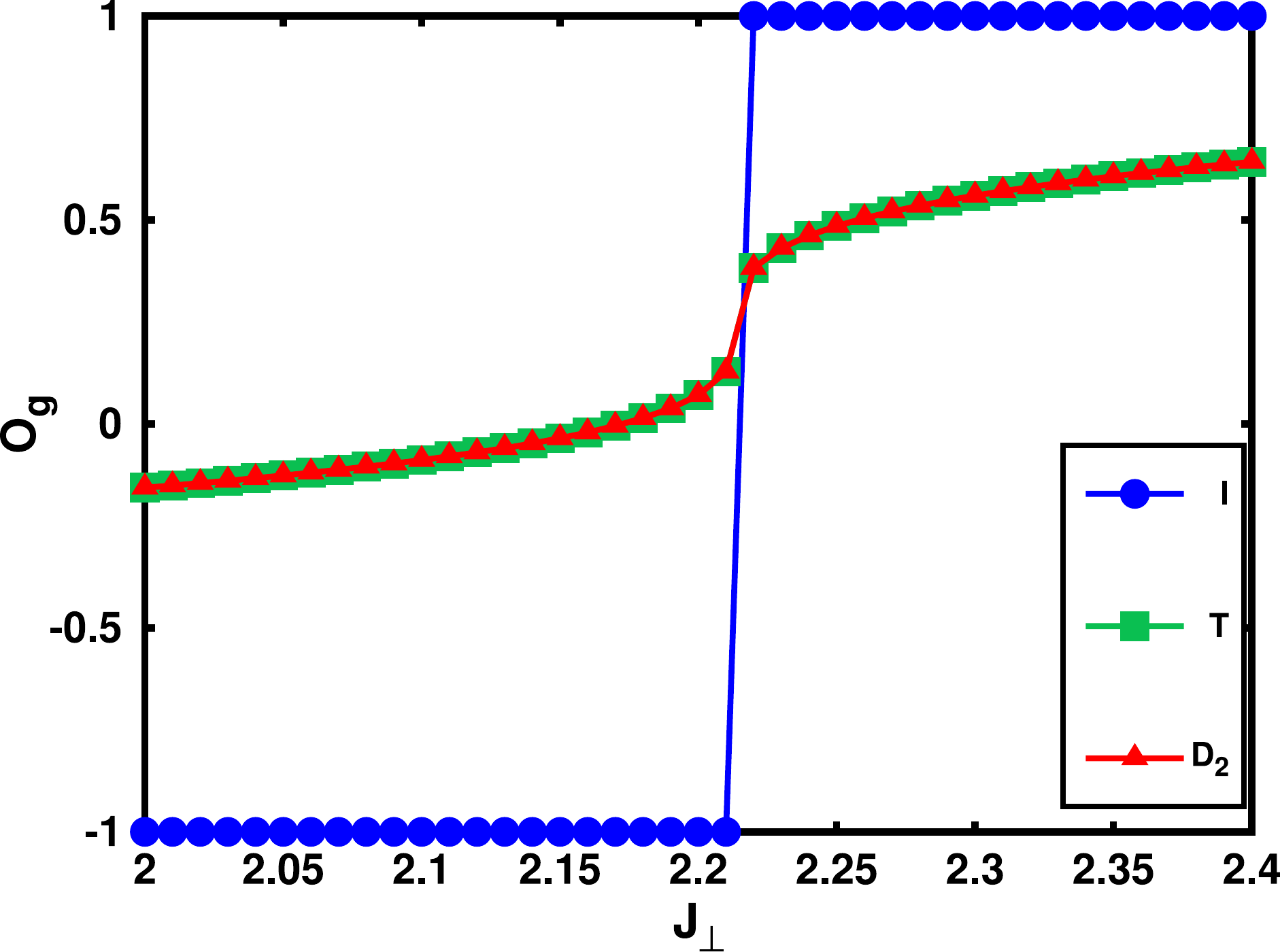}
  \caption{(Colour online) ``Non-local" order parameters $O_g$ for symmetry operations $g = \mathcal{I}$ (blue circle), $g = \mathcal{T}$ (green square) and $g = \mathcal{D}_2$ (red triangle) versus $J_\perp$ with $J_K = 2$ and $U = 0$. The discontinuous change of $O_\mathcal{I}$ from -1 to 1 across $J_\perp^c$ shows the groundstate transits from an SPT phase protected by $\mathcal{I}$ to a topologically trivial phase. The effect of grading on $\mathcal{T}$ and $\mathcal{D}_2$ are apparent in the continuous change of $O_\mathcal{T}$ and $O_{\mathcal{D}_2}$ across $J_\perp^c$.}
  \label{fig:proj-rep}
\end{figure}

Fig. \ref{fig:proj-rep} shows the ``non-local" order parameter $O_g$ corresponding to the three symmetries $\mathcal{I}$, $\mathcal{T}$ and $\mathcal{D}_2$ defined in Eq. (\ref{eqn:non-local-ord-param}) as a function of $J_\perp$. A discontinuous change of $O_\mathcal{I}$ is observed at $J_\perp^c$ where $\mathcal{I}$ changes from a symmetry that protects the SPT phase to one that does not because the groundstate is topologically trivial. The effect of grading of $\mathcal{T}$ and $\mathcal{D}_2$ is obvious in how $O_\mathcal{T}$ and $O_{\mathcal{D}_2}$ change continuously across $J_\perp^c$. However, since increasing $J_\perp$ has the effect of freezing out fermion fluctuations through the formation of local singlets, one can expect that increasing $O_\mathcal{T}$ and $O_{\mathcal{D}_2}$ for $J_\perp > J_\perp^c$ would eventually tend to 1 in the limit of $J_\perp \rightarrow \infty$ since all fermion fluctuations are completely frozen out and the two symmetries $\mathcal{T}$ and $\mathcal{D}_2$ are no longer graded.

\begin{figure}[h!]
  \centering\includegraphics[width=0.45\textwidth]{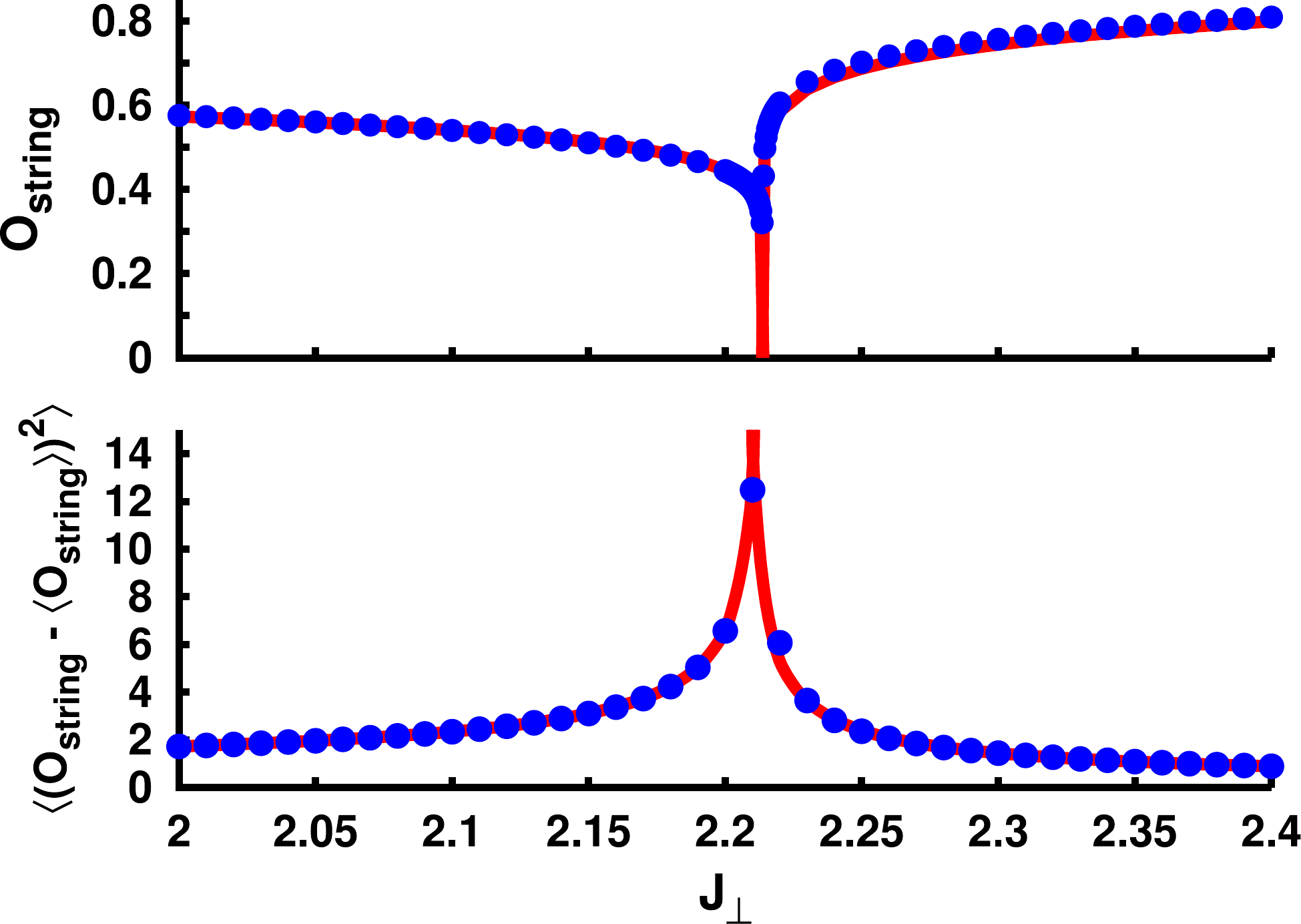}
  \caption{String order parameter $O_\text{string}$ (top) and the variance of $O_\text{string}$ (bottom) versus $J_\perp$ with $J_K = 2$ and $U = 0$. Blue circles are simulation data. Both data are fit with the power law functions $y = a |x - x_c|^\beta$ (top) and $y = b |x - x_c|^\gamma$ (bottom) where $a$, $b$, $\beta$ and $\gamma$ are fitting parameters and $x_c$ is the critical point. The critical point obtained via this fit is $J_\perp^c = 2.2135 \pm 5 \times 10^{-4}$. At $J_\perp = J_\perp^c$, $O_\text{string} = 0$ as the groundstate is highly disordered i.e. there's an equal superposition of all possible fermion site occupation. When $J_\perp > J_\perp^c$, $H_\perp$ dominates and fermions form local Kondo singlets with the local spin-1/2's. This causes fermions to take a one-charge per site ordering which causes $O_\text{string}$ to increase.}
  \label{fig:ord-string}
\end{figure}

The blue circles in the top figure of Fig. \ref{fig:ord-string} show the simulation results of the string order parameter $O_\text{string}$ defined in Eq. (\ref{eqn:string_order}) as a function of $J_\perp$ while the red line is a power law fit of the form $y = a |x - x^c|^\beta$, where $a$ and $\beta$ are fitting parameters and $x_c$ is the critical point. This fit gives the critical point of $J_\perp^c = 2.2135 \pm 5 \times 10^{-4}$ which is within the uncertainty bound of $J_\perp^c$ that obtained from the diverging entropy $S$ in Fig. \ref{fig:svn-U0-JK-Jperp}. The value of $\beta$ gives the critical exponent and it differs slightly when fitting from above and below the critical point: \ignore{$\beta^- = 0.092034$}$\beta^- \approx 9.20 \times 10^{-2}$ and \ignore{$\beta^+ = 0.092684$}$\beta^+ \approx 9.27 \times 10^{-2}$, where the $\beta^-$ ($\beta^+$) is obtained by fitting from below (above) $x_c$. A discontinuity in $O_\text{string}$ occurs at $J_\perp = J_\perp^c$ where the groundstate changes from one with low one particle per site order originating from the non-local fermion hopping of $H_K$, to a groundstate with high one particle per site order due to  $H_\perp$ that tends to suppress fermion hopping by forming local singlets. At exactly $J_\perp = J_\perp^c$, $O_\text{string}$ is expected to vanish, indicating that the state is highly disordered, i.e. each site has an equal distribution of $\hat{n}_l = 0$ and $\hat{n}_l = 1$ and 2. When $J_\perp > J_\perp^c$, $H_\perp$ dominates all other Hamiltonian terms, thus the groundstate is effectively a product of local Kondo singlets and has a large one particle per site order. Increasing $J_\perp$ increases this order and $O_\text{string} \rightarrow 1$ as $J_\perp \rightarrow \infty$. When $J_\perp < J_\perp^c$, the groundstate is dominated by $H_K$ which has a smaller one particle per site ordering than $H_\perp$ due to the non-local fermion hopping $H_K$ contains. The bottom part of Fig. \ref{fig:ord-string} displays the variance of $O_\text{string}$. The variance $\braket{ \left( O_\text{string} - \braket{O_\text{string}} \right)^2}$ is synonymous with the susceptibility $\chi$ since, in analogy to the magnetic susceptibilty $\chi_M$, it measures the fluctuation of the order parameter and diverges at the critical point due to quantum fluctuations. The blue circles in Fig. \ref{fig:ord-string} are the simulation data of the variance of $O_\text{string}$ and the red line is the power law fit $y = b |x - x_c|^\gamma$ where $b$ and $\gamma$ are fit parameters and $x_c$ is the critical point. The value of $\gamma$ is the critical point and its value obtained from fitting from below the critical point is \ignore{$\gamma^- = 0.48556$}$\gamma^- \approx 0.486$, while fitting from above the critical point gives \ignore{$\gamma^+ = 0.65502$}$\gamma^+ \approx 0.655$.

\begin{figure}[h!]
  \centering\includegraphics[width=0.45\textwidth]{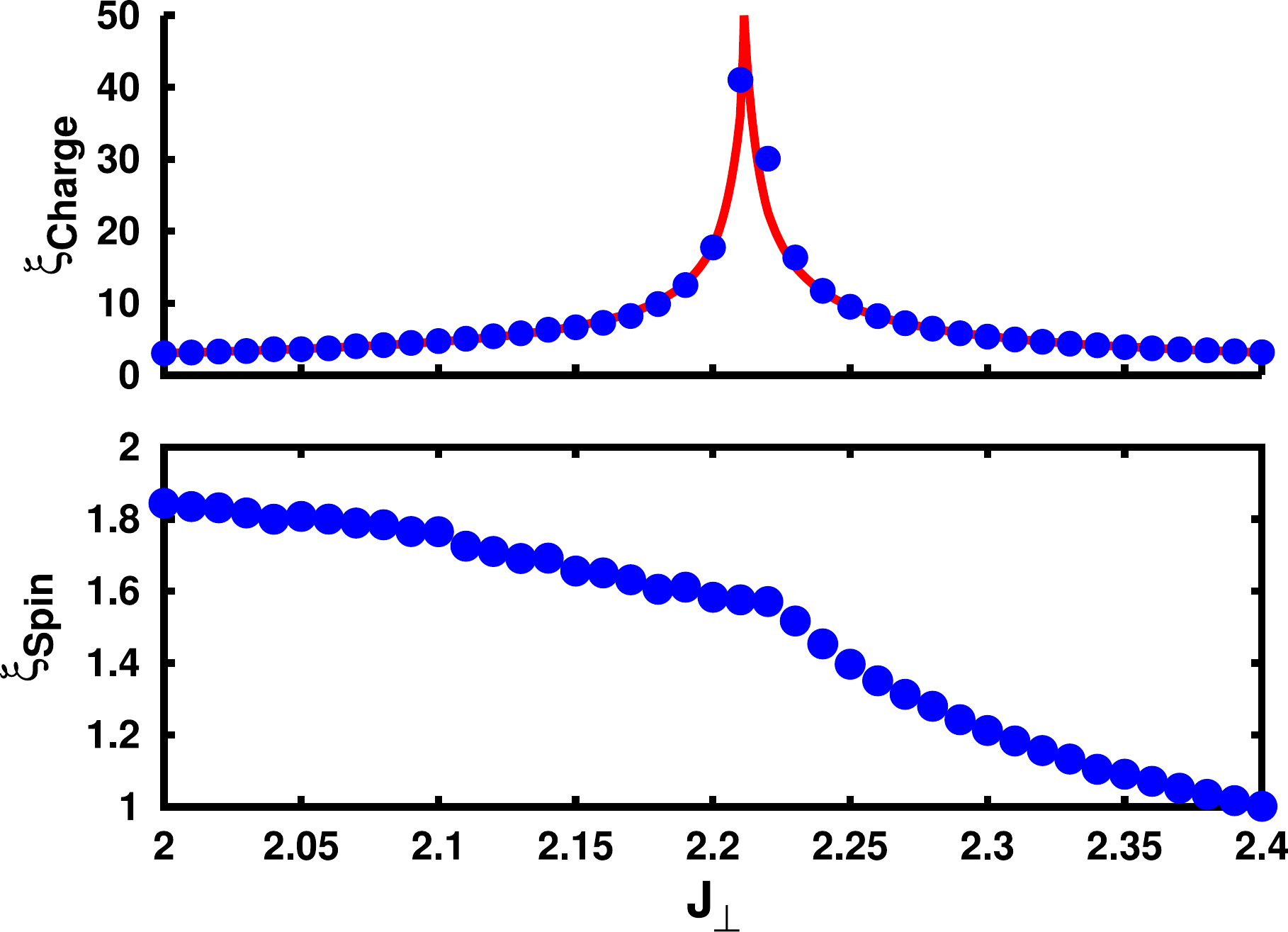}
  \caption{Correlation length of the charge (top) and spin (bottom) excitation versus $J_\perp$ with $J_K = 2$ and $U = 0$. $\xi_\text{Charge}$ diverges at $J_\perp^c$, indicating that the topological phase transition occurs in the charge sector. The red line in the top plot is a power law fit of $y = a |x - x_c |^\nu$ where $a$ and $\nu$ are fitting parameters and $x_c$ is the critical point. The critical point obtained through this fit is $J_\perp^c = 2.21 \pm 0.01$. When $J_\perp < J_\perp^c$, the finite $\xi_\text{Spin}$ is indicative of the Haldane gap while when $J_\perp > J_\perp^c$, $\xi_\text{Spin}$ decreases with $J_\perp$ as local Kondo singlets are formed with increasing $J_\perp$.}
  \label{fig:corr-len}
\end{figure}

In an iMPS, the correlation lengths are computed from the largest eigenvalue of the spectrum of the transfer matrix. These correlation lengths represent any correlation that has the same symmetry as the quantum number of the transfer matrix eigenvalue. Fig. \ref{fig:corr-len} shows $\xi_\text{Charge}$ and $\xi_\text{Spin}$ versus $J_\perp$. These correlation lengths are inversely proportional to the respective energy gaps $\Delta_\text{Charge}$ and $\Delta_\text{Spin}$. At $J_\perp = J_\perp^c$, $\xi_\text{Charge}$ diverges while $\xi_\text{Spin}$ remains non-zero, indicating a topological phase transition in the charge sector. The blue circles are simulation data and the red line in the top plot is a power law fit $y = a |x - x_c|^\nu$ where $a$ and $\nu$ are fitting parameters and $x_c$ is the critical point. The critical point obtained through this fit is $J_\perp^c = 2.21 \pm 0.01$ which within uncertainty bounds agree with the other two values of $J_\perp^c$ obtained in Figs. \ref{fig:svn-U0-JK-Jperp} and \ref{fig:ord-string}. The value of $\nu$ gives the critical exponent and it differs slightly when fitting from above and below the critical point: \ignore{$\nu^- = 0.66643$}$\nu^- \approx 0.666$ and \ignore{$\nu^+ = 0.74169$}$\nu^+ \approx 0.742$, where the $\nu^-$ ($\nu^+$) is obtained by fitting from below (above) $x_c$. When $J_\perp < J_\perp^c$, the local triplets formed between fermions and local spin-1/2's mimic a Heisenberg spin-1 chain which is known to be an insulating Haldane groundstate containing a non-zero spin gap $\Delta_\text{Spin} \propto 1 / \xi_\text{Spin}$. When $J_\perp > J_\perp^c$, the groundstate forms local singlets and $\xi_\text{spin}$ decreases with increasing $J_\perp$ as the system now is a trivial product of local singlets where the binding energy (or energy gap) of the singlets increases with $J_\perp$.

\begin{figure}[h!]
  \centering\includegraphics[width=0.45\textwidth]{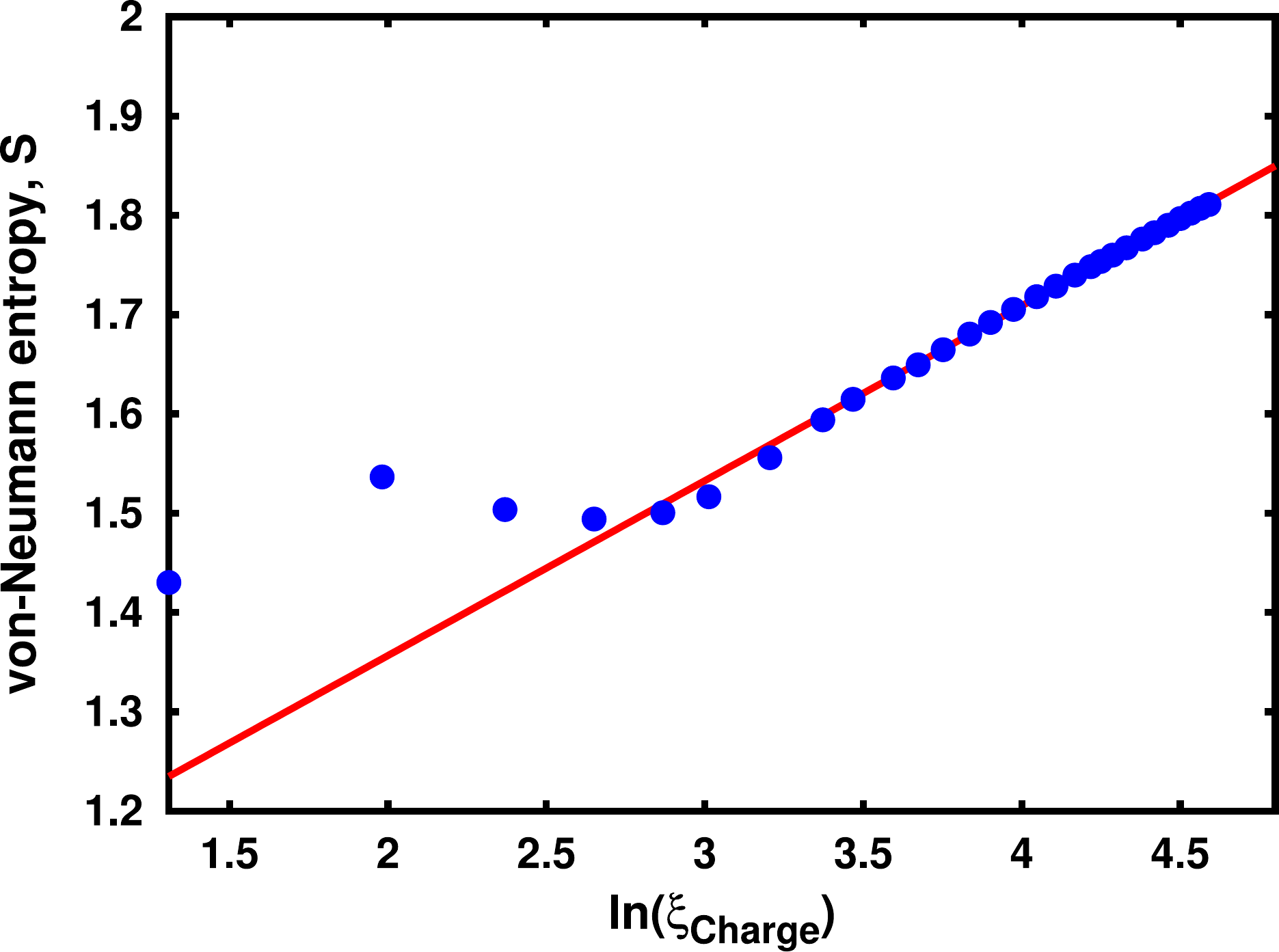}
  \caption{von-Neumann entropy $S$ versus $\ln \xi_\text{Charge}$ for $J_\perp = J_\perp^c = 2.2140$. Blue circles are simulation data while the solid red line is the linear fit $y = a \ln \xi_\text{Charge} + b$, where $a$ and $b$ are fitting parameters with values \ignore{$0.17613$}$0.176$ and \ignore{$1.0043$}$1.00$ respectively. The value of $a$ gives the central charge $c = 6a \approx 1.06$.}
  \label{fig:svn-logxi}
\end{figure}

For a critical 1D system, the von-Neumann entropy is known to scale logarithmically with the correlation length $\xi$ according to
\begin{eqnarray}
S = \frac{c}{6} \ln \frac{\xi}{l},
\label{eqn:svn_xi_relation}
\end{eqnarray}
where $c$ is the central charge, which is the number of degrees of freedom of the system that are critical. $\xi$ is the correlation length and $l$ is a short-distance length scale (e.g. lattice spacing) \cite{Calabrese}. Fig. \ref{fig:svn-logxi} shows the von-Neumann entropy versus $\ln \xi_\text{Charge}$ at $J_\perp^c$. The linear fit (red line) of $y = a \ln \xi_\text{Charge} + b$ in this figure has a gradient \ignore{$a = 0.17613$}$a \approx 0.176$, which upon comparing to Eq. (\ref{eqn:svn_xi_relation}) gives a central charge $c = 6a \approx 1.06$. This value of the central charge is close to that of the 1D isotropic quantum Heisenberg ($XXX$) model and the 1D free fermions/bosons model, both having a central charge $c = 1$. Since the central charge tells the number of degrees of freedom that is critical, the value of $c \approx 1$ together with the diverging charge excitation correlation length $\xi_\text{Charge}$ corroborates that the spin excitation correlation length must not diverge and the spin excitation sector, $\xi_\text{Spin}$ is not critical. The deviation of $S$ from the linear fit in Fig. \ref{fig:svn-logxi} at small $\ln (\xi)$ is due to the fact that at a small basis size $m$, the wavefunction is not a good representation of the actual groundstate.

\begin{figure}[h!]
  \centering\includegraphics[width=0.45\textwidth]{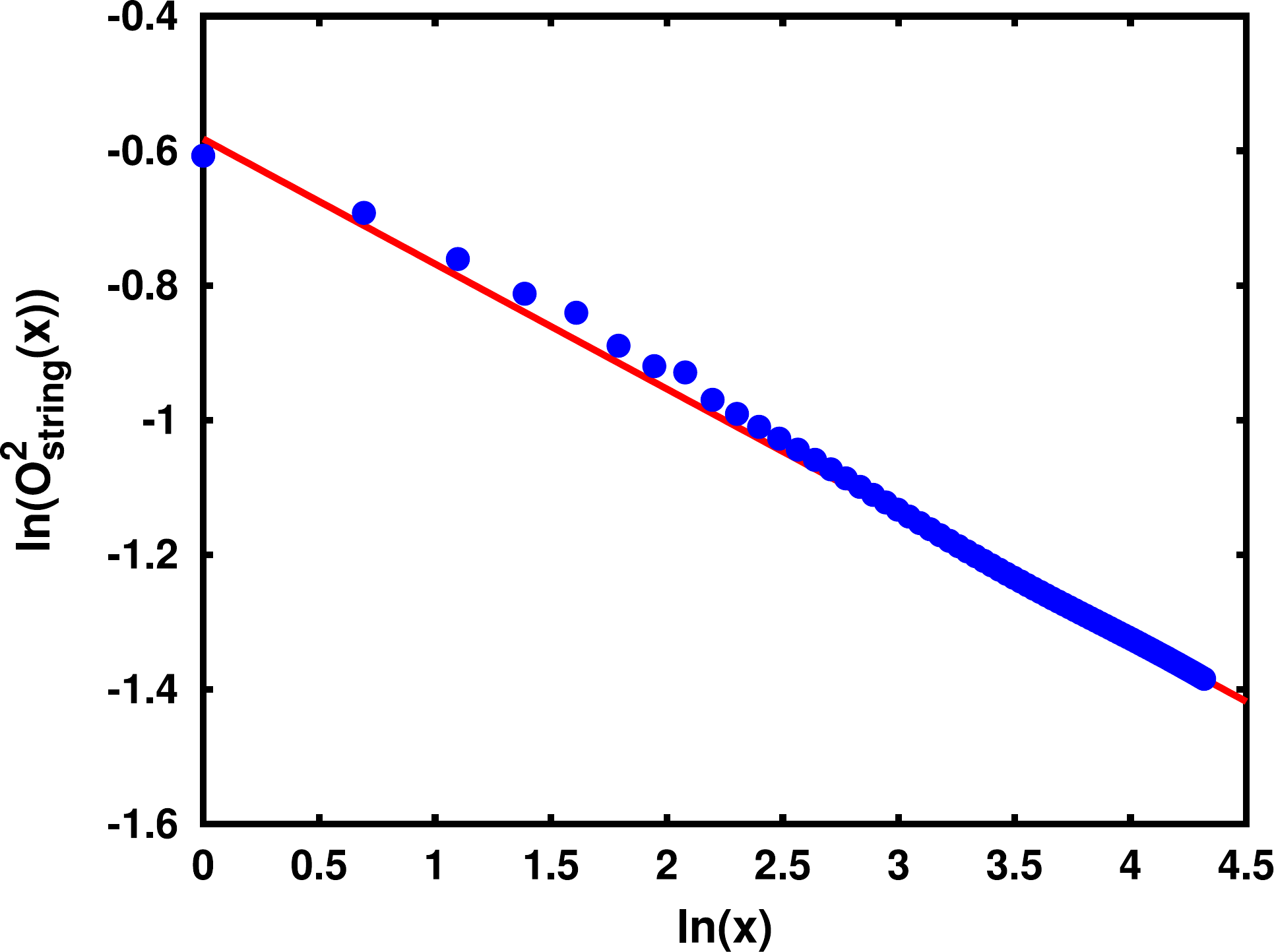}
  \caption{$\log$-$\log$ plot of the correlation function $O_\text{string}^2(x)$ vs lattice position $x$ for $J_\perp = J_\perp^c = 2.2140$. Blue circles are simulation data while the solid red line is the linear fit $y = a \ln x + c$, where $a$ and $c$ are fitting parameters with values \ignore{$-0.18569$}$-0.186$ and \ignore{$-0.58234$}$-0.582$ respectively. The fitting parameter $a$ is used obtain the critical exponent \ignore{$\eta = 1 - a = 1.1857$}$\eta = 1 - a \approx 1.19$.}
  \label{fig:log-ord-x-log-x}
\end{figure}

At the critical point, the exponents extracted from physical observables such as the order parameter, susceptibility, correlation function, etc, are known to obey scaling identities that relate the different exponents to each other. Some well known scaling identities include
\begin{eqnarray}
\alpha + 2\beta + \gamma &=& 2 \quad \text{(Rushbrooke's identity)} , \nonumber
\\
\delta - 1 &=& \frac{\gamma}{\beta} \quad \text{(Widom's identity)} , \nonumber
\\
(2 - \eta) \nu &=& \gamma \quad \text{(Fisher's identity)} , \nonumber
\\
2 - \alpha &=& \nu d \quad \text{(Josephson's identity)}.
\label{eqn:scaling_id}
\end{eqnarray}
The last expression involving the spatial dimension $d$ is also known as the hyperscaling relation. To show that the critical exponents obtained above obey the scaling relation, another critical exponent is extracted from the string order parameter $O_\text{string}^2$ by using it as a string correlation function,
\begin{eqnarray}
O_\text{string}^2(x) = \braket{p(0) p(x)}
\label{eqn:string_order_corr}
\end{eqnarray}
where $p(x) \equiv \prod_{j < x} (-1)^\frac{n_j - 1}{2}$ is a ``kink" operator that measures the 2-particle fluctuation with net zero spin at point $x$ on the lattice. Eq. (\ref{eqn:string_order_corr}) is equal to $O_\text{string}^2$ defined Eq. (\ref{eqn:string_order}) and serves only to show that $O_\text{string}^2$ acts similar to a 2-point correlation function. The details of the conversion of Eq. (\ref{eqn:string_order}) into Eq. (\ref{eqn:string_order_corr}) is presented in Section \ref{appendix_Ostring}. The key difference between $O_\text{string}^2$ that is used as an order parameter and $O_\text{string}^2(x)$ that is used as a correlation function is that in the former, the spatial points are taken to infinity as shown in the limit $j - k \rightarrow \infty$ in Eq. (\ref{eqn:string_order}), whereas the latter is computed only on a finite region of the lattice. Fig. \ref{fig:log-ord-x-log-x} displays $O_\text{string}^2(x)$ as a function of lattice position $x$ on a $\log$-$\log$ scale, at the critical point $J_\perp^c$. Conventionally, the correlation function of choice for this is the two-point density function $\braket{\rho(x) \rho(x')}$. However, since it is the particle number fluctuation $O_\text{string}$ that correctly captures the phase transition, the spatial correlation $O_\text{string}(x)$ is chosen over $\braket{\rho(x) \rho(x')}$. In addition to this, it was found that $O_\text{string}(x)$ decayed much slower than the density correlation function (not shown here), hence $O_\text{string}(x)$ serves as a preferred choice of order parameter. The blue circle are the simulation data while the solid red line is the linear fit $y = a \ln x + c$ where $a$ and $c$ are fitting parameters. $m$ is the gradient and it is related to the critical exponent $\eta$ via the relation $a = 2 - d - \eta$, where $d$ is the spatial dimension and is taken to be unity in this 1D model. The value of $a$ obtained from the linear fit is \ignore{$a = -0.18569$}$a = -0.186$, which gives $\eta = 1 - a \approx 1.19$.

The four critical exponents $\beta^\pm$ (order parameter $O_\text{string}$), $\nu^\pm$ (Charge correlation length $\xi_\text{Charge}$), $\gamma^\pm$ (variance of $O_\text{string}$) and $\eta$ (correlation function $O_\text{string}(x)$) can now be used to check the scaling relations. For consistency, Fisher's identity $\gamma'^\pm = (2 - \eta) \nu^\pm$ can first be used to compare the values of $\gamma^\pm$:
\begin{eqnarray}
\gamma'^- = (2 - \eta) \nu^- = (2 - 1.19) \times 0.666 = 0.54 , \nonumber
\\
\gamma'^+ = (2 - \eta) \nu^+ = (2 - 1.19) \times 0.742 = 0.60 , \nonumber
\end{eqnarray}
which differ from $\gamma^-$ and $\gamma^+$ by $11\%$ and $8\%$ respectively. Alternatively, Rushbrooke's identity can be inserted into Josephson's identity to eliminate $\alpha$ and give the relation $\gamma'^\pm = \nu^\pm d - 2\beta^\pm$. Using $d = 1$, the values of $\gamma^\pm$ can again be compared:
\begin{eqnarray}
\gamma'^- = \nu^- - 2\beta^- = 0.666 - 2 \times 0.092 = 0.482 , \nonumber
\\
\gamma'^+ = \nu^+ - 2\beta^+ = 0.742 - 2 \times 0.0927 = 0.556 , \nonumber
\end{eqnarray}
which differ from $\gamma^-$ and $\gamma^+$ by $0.8\%$ and $15\%$ respectively. With this consistency check, it is now possible to use the four scaling identities and the four known exponents $\beta$, $\gamma$, $\nu$ and $\eta$ to obtain the unknown exponents $\alpha$ and $\delta$. There are three ways that the scaling identities can be manipulated to give $\alpha$:
\begin{eqnarray}
\alpha_R &=& 2(1 - \beta) - \gamma \text{ (Rushbrooke)} , \nonumber
\\
\alpha_J &=& 2 - \nu d \text{ (Josephson)} , \nonumber
\\
\alpha_{R-F} &=& 2(1 - \beta) - (2 - \eta)\nu \text{ (Rushbrooke-Fisher)}, \nonumber
\end{eqnarray}
where the third identity is obtained by eliminating $\gamma$ by substituting the Rushbrooke identity into the Fisher identity. As for $\delta$, there are two ways to obtain it:
\begin{eqnarray}
\delta_W &=& \frac{\gamma}{\beta} - 1 \text{ (Widom)} , \nonumber
\\
\delta_{W-F} &=& \frac{(2 - \eta)\nu}{\beta} - 1 \text{ (Widom-Fisher)} , \nonumber
\end{eqnarray}
where the second identity is obtained by relating the Fisher identity to Widom's identity through $\gamma$. The subscript of $\alpha$'s and $\delta$'s above are labels that show the identities that they calculated from. The different ways of obtaining $\alpha$ and $\delta$ give a slight variation in their values through the variation to the different known exponents $\beta$, $\gamma$, $\nu$ and $\eta$. These values of $\alpha$ and $\delta$ are tabulated together with the known exponents, and their analogous, more familiar physical observables (e.g. specific heat, magnetization, etc.) in Table \ref{table:critical-exp}. The exponent $\delta$ relates the order parameter $O_{\text{string}}$ to the source field $J$ (analagous to the magnetic field $H$ for the Curie point, or the scaled pressure $\frac{P - P_c}{P_c}$ in the liquid-gas transition), however it is interesting to note that it is not
obvious how to construct an explicit form for the $J$ field corresponding to a string order parameter.

\begin{table*}
\centering
\begin{tabularx}{\textwidth}{|Y|Y|Y|Y|Y|Y|Y|}
\hline
Critical exponent & $\alpha$ & $\beta$ & $\gamma$ & $\delta$ & $\nu$ & $\eta$ \\
\hline
Calculated observable/scaling relation & Scaling identity $\alpha_R = 2(1 - \beta) - \gamma$, $\alpha_J = 2 - \nu d$, $\alpha_{R-F} = 2(1 - \beta) - (2 - \eta)\nu$ & String order parameter $O_\text{string}$ & Variance of $O_\text{string}$ & Scaling identity $\delta_W = \frac{\gamma}{\beta} - 1$, $\delta_{W-F} = \frac{(2 - \eta)\nu}{\beta} - 1$ & Correlation length of charge excitation, $\xi_\text{Charge}$ & String correlation function $O_\text{string}(x)$ \\
\hline
Analogous known physical observable ($\tau \equiv \frac{T - T_c}{T_c}$) & Specific heat $C \propto \tau^{-\alpha}$ & Order parameter (e.g. Magnetization for the Curie point) $\Psi \propto \tau^\beta$ & Susceptibility ($\frac{d \psi}{dJ}$), $\chi \propto \tau^\gamma$ & Source field (e.g. Magnetic field $H$ for the Curie point) $J \propto \Psi^\delta$ & Correlation length $\xi \propto \tau^{-\nu}$ & Correlation function $\braket{\psi(0) \psi(x)} \propto x^{2 - d - \eta}$ \\
\hline
Value & $\alpha_R^- = 1.33$, $\alpha_R^+ = 1.16$, $\alpha_J^- = 1.33$, $\alpha_J^+ = 1.26$, $\alpha_{R-F}^- = 1.27$, $\alpha_{R-F}^+ = 1.21$ & $\beta^- = 9.20 \times 10^{-2}$, $\beta^+ = 9.27 \times 10^{-2}$ & $\gamma^- = 0.486$, $\gamma^+ = 0.655$ & $\delta_W^- = 6.28$, $\delta_W^+ = 8.07$, $\delta_{W-F}^- = 6.90$, $\delta_{W-F}^+ = 7.52$ & $\nu^- = 0.666$, $\nu^+ = 0.742$ & $\eta = 1.19$ \\
\hline
\end{tabularx}
\caption{List of critical exponents. The first row is the critical exponents. The second row shows the observable or scaling relation used to obtain the critical exponents in this work. The third row shows more familiar physical observables used in statistical mechanics that are analogous to the observables used in this work to obtain the critical exponents. The fourth row displays the values of the critical exponents obtained in this work.}
\label{table:critical-exp}
\end{table*}

%%%%%%%%%%%%%%%%%%%%%%%%%%%%%%%%%%%%%%%%%%%%%%%%%%%%%%%%%%%%%%%%%%%%%%%%%%%%%%%%%%%%%%%%%%%%%%%%%%%%%%%%%%%
%%%%%%%%%%%%%%%%%%%%%%%%%%%%%%%%%%%%%%%%%%%%%%%%%%%%%%%%%%%%%%%%%%%%%%%%%%%%%%%%%%%%%%%%%%%%%%%%%%%%%%%%%%%

\subsection{Topological phase transition with $U > 0$}
\label{Coulomb-svn3}
\begin{figure}[h!]
  \centering\includegraphics[width=0.45\textwidth]{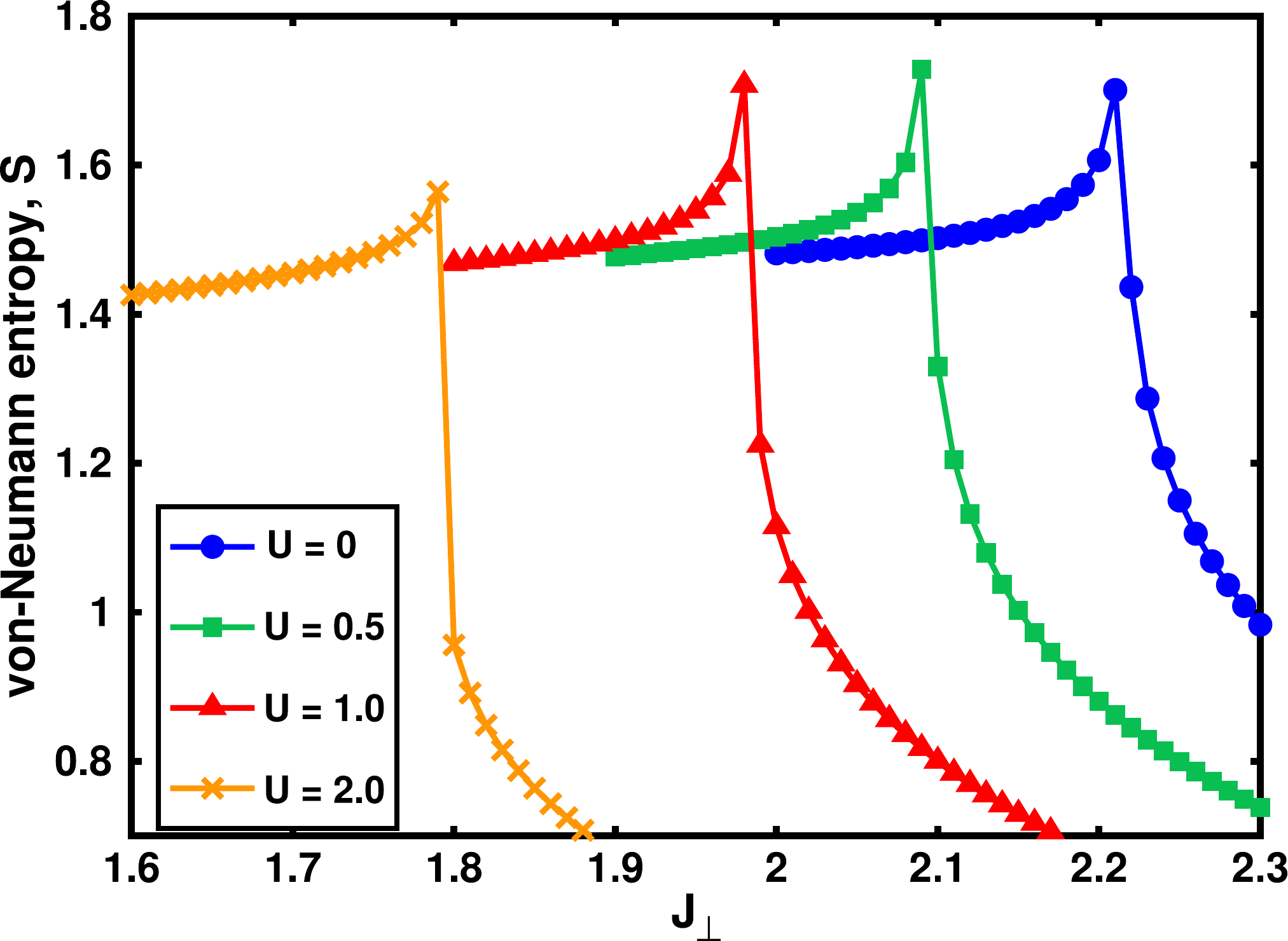}
  \caption{(Colour online) von-Neumann entropy $S$ versus $J_\perp$ with $J_K = 2$, for $U = $ 0, 0.5, 1 and 2. Increasing $U$ decreases the value of $J_\perp$ where the critical point occurs due to the effect of Hubbard interaction that tends to form a one-charge per site order, thus assisting in the formation of local singlets.}
  \label{fig:svn-U-JK-Jperp}
\end{figure}

\begin{figure}[h!]
  \centering\includegraphics[width=0.45\textwidth]{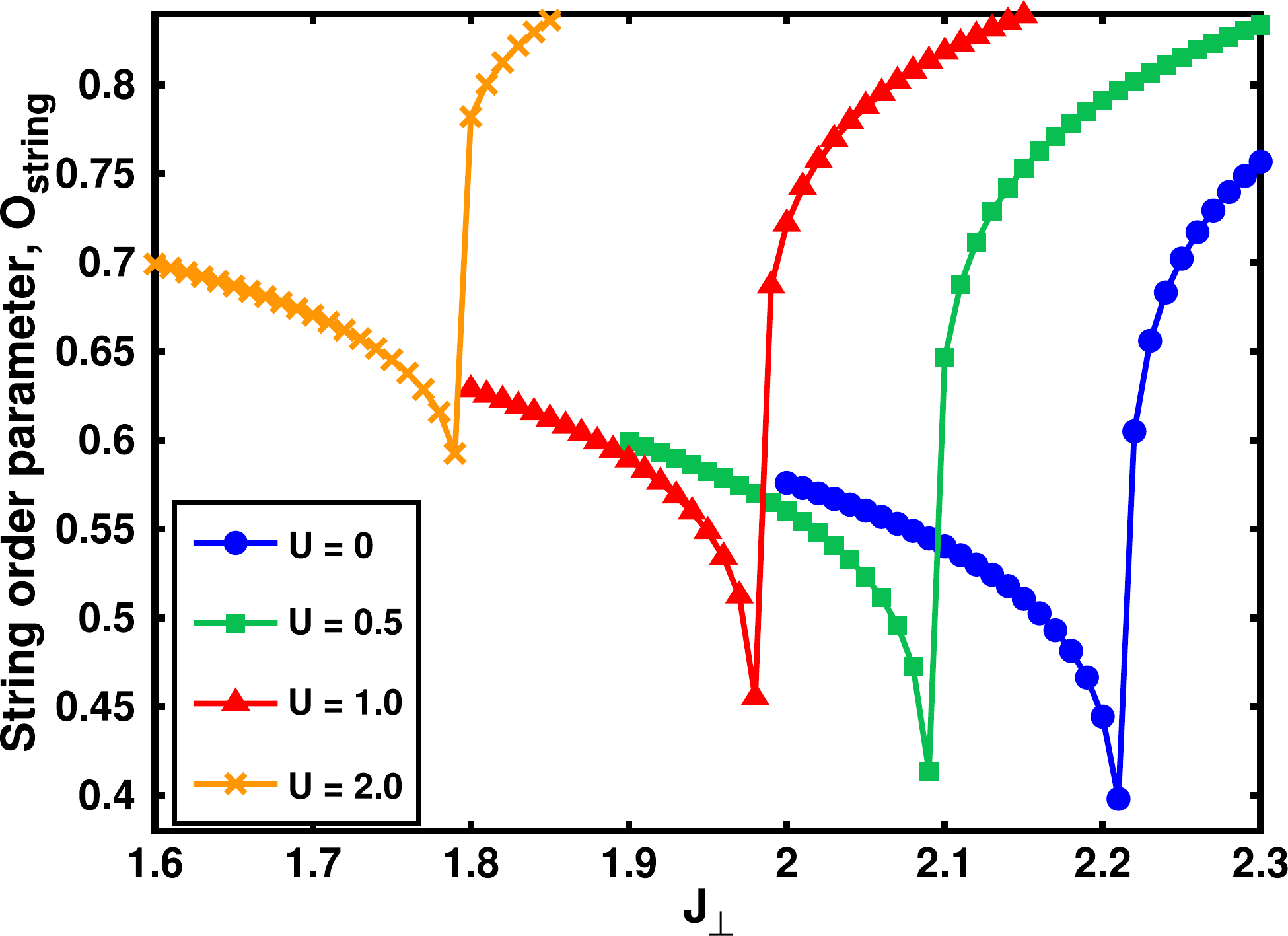}
  \caption{(Colour online) $O_\text{string}$ versus $J_\perp$ with $J_K = 2$, for $U = $ 0, 0.5, 1 and 2. The formation of a one-charge per site order increases with $U$ since fermions repel each other more strongly, thus lowering the value of $J_\perp^c$ required to transform the groundstate to from one favoring non-local fermion hopping and low one-charge per site order, to one that is a direct product of local singlets with large one-charge per site occupation.}
  \label{fig:ord-string-U-JK-Jperp}
\end{figure}

In this section, $J_K$ is fixed at 2 while $U$ and $J_\perp$ are varied. By tuning $U > 0$, the critical point $J_\perp^c$ shifts to smaller values as shown in Figs. \ref{fig:svn-U-JK-Jperp} and \ref{fig:ord-string-U-JK-Jperp}. As explained earlier in Section \ref{Coulomb-svn1}, this occurs because both the Hubbard interaction in the Hubbard chain described by Eq. (\ref{eqn_Hamiltonian_Hubbard}), and the local $s$-wave Kondo coupling between the Heisenberg and Hubbard chains $H_\perp$ described by Eq. (\ref{eqn_Hamiltonian_Kondo_local}) favor a groundstate containing a one fermion per-site order compared to an empty or doubly occupied site order. Therefore increasing $U$ while keeping $J_K$ fixed, reduces the effect of the non-local hopping in $H_K$, making it easier for $H_\perp$ to form local singlet and hence reducing $J_\perp^c$.

\begin{figure}[h!]
  \centering\includegraphics[width=0.45\textwidth]{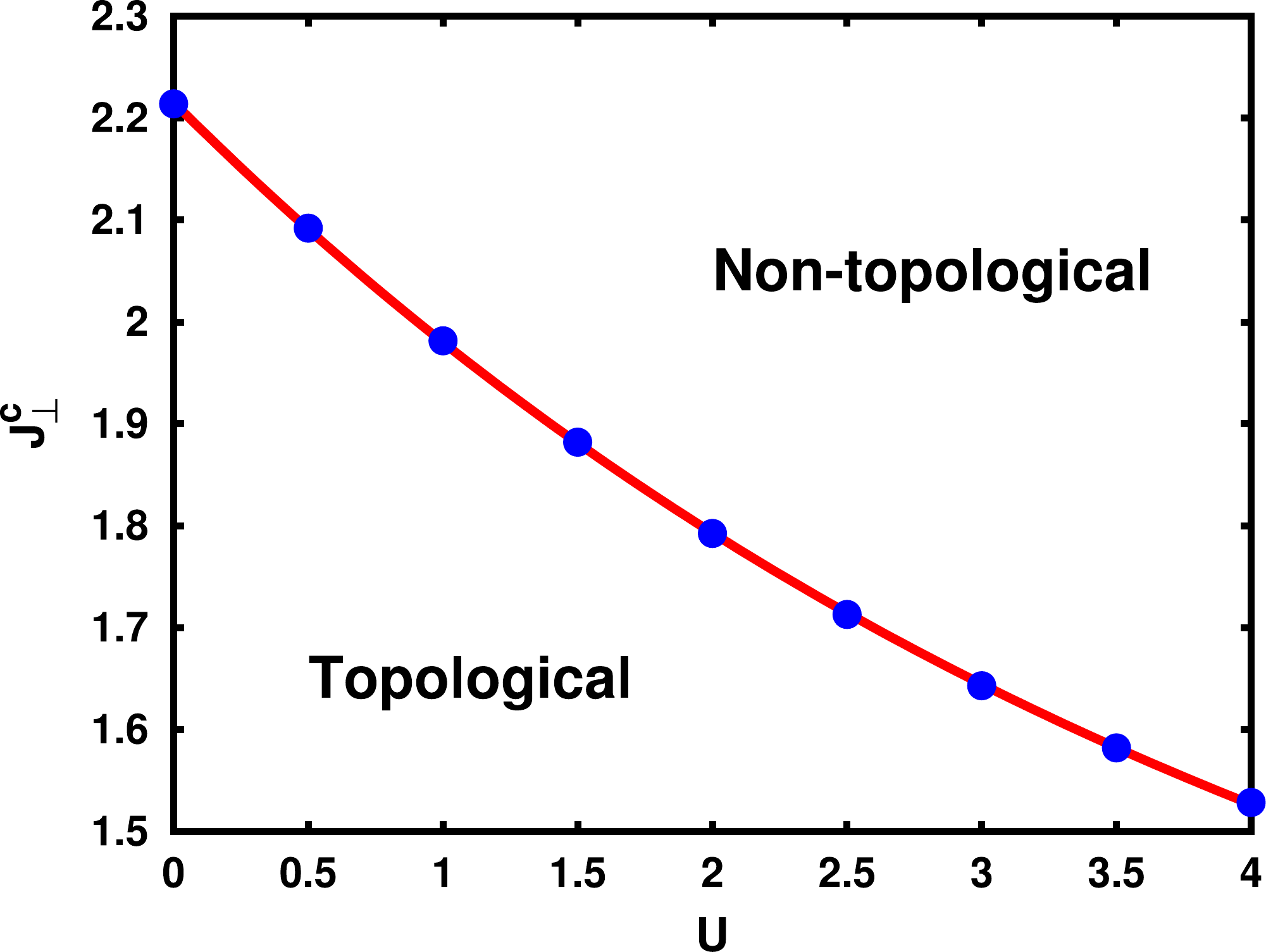}
  \caption{Phase diagram separating the SPT phase from the topologically trivial phase with $J_K = 2$. Blue circles are data points while the red line is the exponential fit of the form $y = ae^{-bx} + c$, where $a$, $b$ and $c$ are fitting parameters. Error of each data point is of the order of $10^{-4} - 10^{-5}$.}
  \label{fig:phase-diagram}
\end{figure}

By plotting the value of $J_\perp^c$ for different values of $U$, one obtains a phase diagram of $J_\perp^c$ against $U$, separating the SPT phase from the topologically trivial phase as shown in Fig. \ref{fig:phase-diagram}. The red line in Fig. \ref{fig:phase-diagram} is an exponential fit of the form $y = ae^{-bx} + c$, where $a$, $b$ and $c$ are fitting parameters. Since an exponential decay function only vanishes in the limit of $x \rightarrow \infty$ and $c = 0$, this fit indicates that even for large, finite values of $U$, the SPT phase survives but is easily destroyed by small $J_\perp$. This statement is only valid within the range of $U = 0 - 4$ of the phase diagram Fig. \ref{fig:phase-diagram} since the wavefunctions of the simulations become non-injective when $U > 5$ due to the SU(2) symmetry enforced on the wavefunctions. This non-injectivity of the wavefunction causes the Schmidt values to be degenerate and any quantity that depends on it becomes unreliable across a topological phase transition. One can however qualitatively guess how the phase diagram may look like when $U > 4$ by gathering information from Fig. \ref{fig:svn-U-J_K} when $U > 4$ and comparing it to the diverging $S$'s in Fig. \ref{fig:svn-U-JK-Jperp}. Up to $U = 10$, $S$ in Fig. \ref{fig:svn-U-J_K} decays exponentially. Thus, if $H_\perp$ was now introduced, $J_\perp^c$ would occur at exponentially decreasing values as $U$ is increased up to $U = 10$. This is consistent with the results in the phase diagram Fig. \ref{fig:phase-diagram}. When $U > 10$, $S$ in Fig. \ref{fig:svn-U-J_K} plateaus with increasing $U$. Thus if $H_\perp$ is now introduced, $J_\perp^c$ too would even out with increasing $U$. Therefore, one would qualitatively expect that the phase diagram of $J_\perp^c$ versus $U$ would be relatively flat when $U > 10$.

The precise value of $J_\perp^c$ becomes more difficult to obtain with increasing $U$ as the peaks of $S$ become narrower. In order to obtain $J_\perp^c$ precisely requires a finer increment of $J_\perp$. The finest increment of $J_\perp$ done in this work is $1 \times 10^{-5}$ and this is found to be insufficient to precisely obtain $J_\perp^c$ for the purpose of finding the central charge when $U > 1$. Since $S$ near $J_\perp^c$, but not precisely at $J_\perp^c$, does not scale according to Eq. (\ref{eqn:svn_xi_relation}), the calculations of the central charge for different non-zero $U$'s was not satisfactory and will be reserved for future work. Regardless of this, there are no \textit{a priori} reasons to expect that the central charge will be affected by $U$.

%%%%%%%%%%%%%%%%%%%%%%%%%%%%%%%%%%%%%%%%%%%%%%%%%%%%%%%%%%%%%%%%%%%%%%%%%%%%%%%%%%%%%%%%%%%%%%%%%%%%%%%%%%%
%%%%%%%%%%%%%%%%%%%%%%%%%%%%%%%%%%%%%%%%%%%%%%%%%%%%%%%%%%%%%%%%%%%%%%%%%%%%%%%%%%%%%%%%%%%%%%%%%%%%%%%%%%%

\section{Summary}
The groundstate of the TKI is shown to be in a Haldane phase protected by inversion symmetry. While the effect of Hubbard interaction is to create a single fermion per site order, it by itself is insufficient to destroy the SPT phase caused by the non-local fermion hopping originating from the $p$-wave coupling. When the conventional $s$-wave Kondo coupling is introduced, it competes with the $p$-wave coupling by suppressing the non-local fermion hopping through the formation of local singlets. This causes a topological phase transition from a SPT phase to a topologically trivial phase when $J_\perp > J_\perp^c$ which is evident in the von-Neumann entropy, ES, $O_\text{string}$ and ``non-local" order parameters defined in Eq. (\ref{eqn:non-local-ord-param}). The topological phase transition occurs only in the charge sector, where the correlation length of the charge excitation $\xi_\text{Charge}$ diverges, while the correlation length of the spin excitation $\xi_\text{Spin}$ remains finite. Such a transition is rather unusual, and indicates that there is an effective pairing which causes a phase transition driven by (spinless) two-particle excitations, while the single-particle gap remains non-zero. The origin of this pairing is not obvious from the bare interactions in the Hamiltonian. At the critical point, the critical exponents extracted from the order parameter $O_\text{string}$ and the correlation length of the charge sector $\xi_\text{Charge}$ fit the scaling relations. The central charge $c \approx 1$ obtained from the von-Neumann entropy $S$ shows the that the transition belongs to the same universality class as 1D free bosons. The effect of forming local singlets is enhanced when the Hubbard interaction and $s$-wave coupling are introduced simultaneously, thus reducing $J_\perp^c$ when $U$ is increased.

%%%%%%%%%%%%%%%%%%%%%%%%%%%%%%%%%%%%%%%%%%%%%%%%%%%%%%%%%%%%%%%%%%%%%%%%%%%%%%%%%%%%%%%%%%%%%%%%%%%%%%%%%%%
%%%%%%%%%%%%%%%%%%%%%%%%%%%%%%%%%%%%%%%%%%%%%%%%%%%%%%%%%%%%%%%%%%%%%%%%%%%%%%%%%%%%%%%%%%%%%%%%%%%%%%%%%%%

\section{Acknowledgement}
The authors thank Alejandro Mezio, Henry L. Nourse, and S. Nariman Saadatmand for helpful discussions. This work has been supported by the Australian Research Council (ARC) Centre of Excellence for Engineered Quantum Systems, Grant No. CE110001013. I.P.M. also acknowledges support from the ARC Future Fellowships Scheme No. FT140100625.

%%%%%%%%%%%%%%%%%%%%%%%%%%%%%%%%%%%%%%%%%%%%%%%%%%%%%%%%%%%%%%%%%%%%%%%%%%%%%%%%%%%%%%%%%%%%%%%%%%%%%%%%%%%
%%%%%%%%%%%%%%%%%%%%%%%%%%%%%%%%%%%%%%%%%%%%%%%%%%%%%%%%%%%%%%%%%%%%%%%%%%%%%%%%%%%%%%%%%%%%%%%%%%%%%%%%%%%

%%%%%%%%%%%%%%%%%%%%%%%%%%%%%%%%%%%%%%%%%%%%%%%%%%%%%%%%%%%%%%%%%%%%%%%%%%%%%%%%%%%%%%%%%%%%%%%%%%%%%%%%%%%
%%%%%%%%%%%%%%%%%%%%%%%%%%%%%%%%%%%%%%%%%%%%%%%%%%%%%%%%%%%%%%%%%%%%%%%%%%%%%%%%%%%%%%%%%%%%%%%%%%%%%%%%%%%

\section{Appendix}
\subsection{Scaling of correlation length $\xi$ and von-Neumann entropy $S$ with respect to basis size $m$}
\label{appendix_scaling}
\begin{figure}[h!]
  \centering\includegraphics[width=0.45\textwidth]{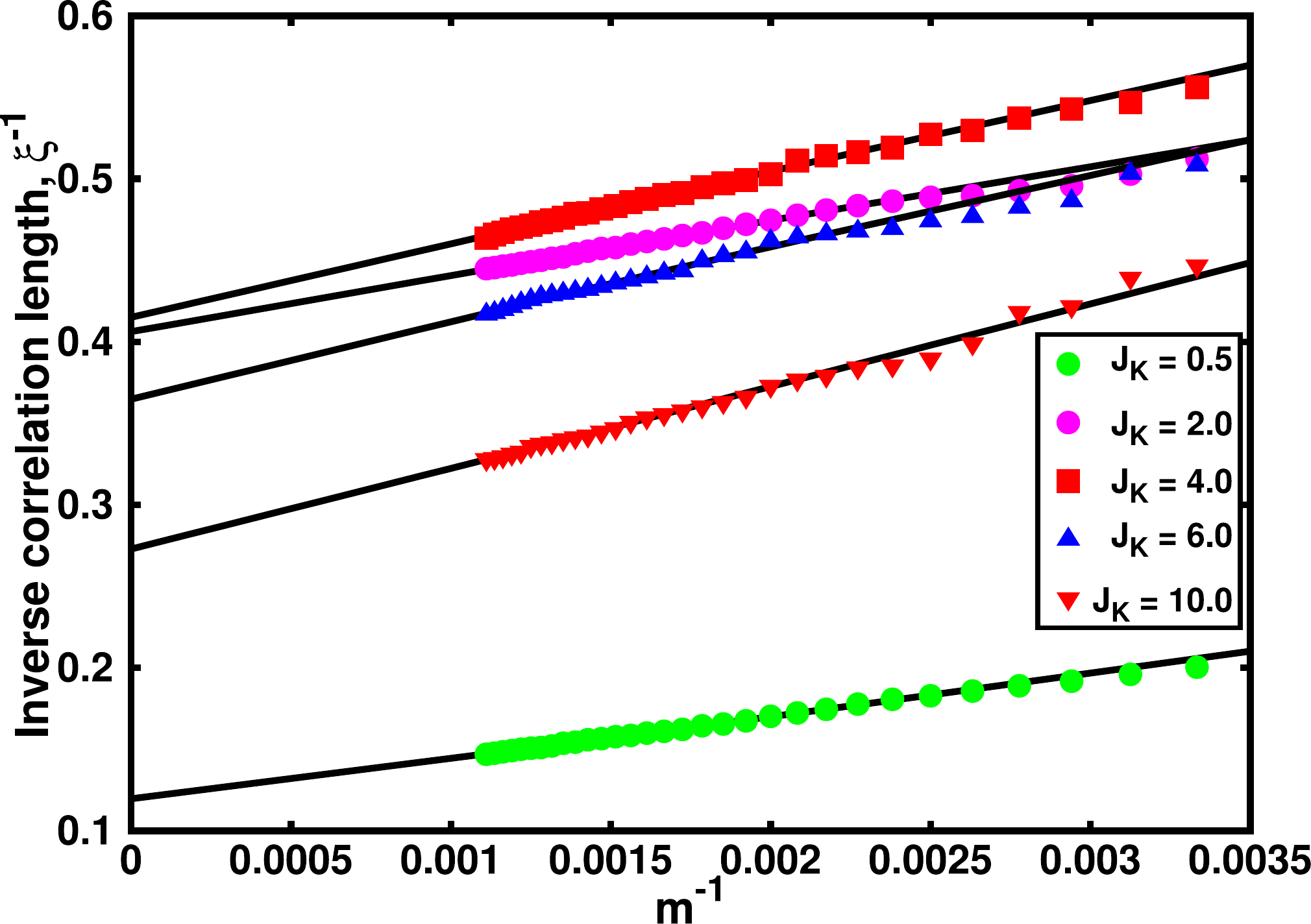}
  \caption{(Colour online) Inverse correlation length $\xi^{-1}$ of spin excitation versus basis size $m^{-1}$. Symbols are simulation data for different values of $J_K$ shown in the legend. For clarity and to avoid clutter, not all values of $J_K$ are shown here. Black lines are the fit of form $\xi^{-1} = a m^{-\kappa} + \xi_0^{-1}$, where $a$, $\kappa$ and $\xi_0^{-1}$ are fitting parameters. $\xi_0^{-1}$ is the $y$-axis intercept and represent the value of $\xi_0^{-1}$ as $m \rightarrow \infty$.}
  \label{fig:inverse_corr_len_vs_m}
\end{figure}

\begin{figure}[h!]
  \centering\includegraphics[width=0.45\textwidth]{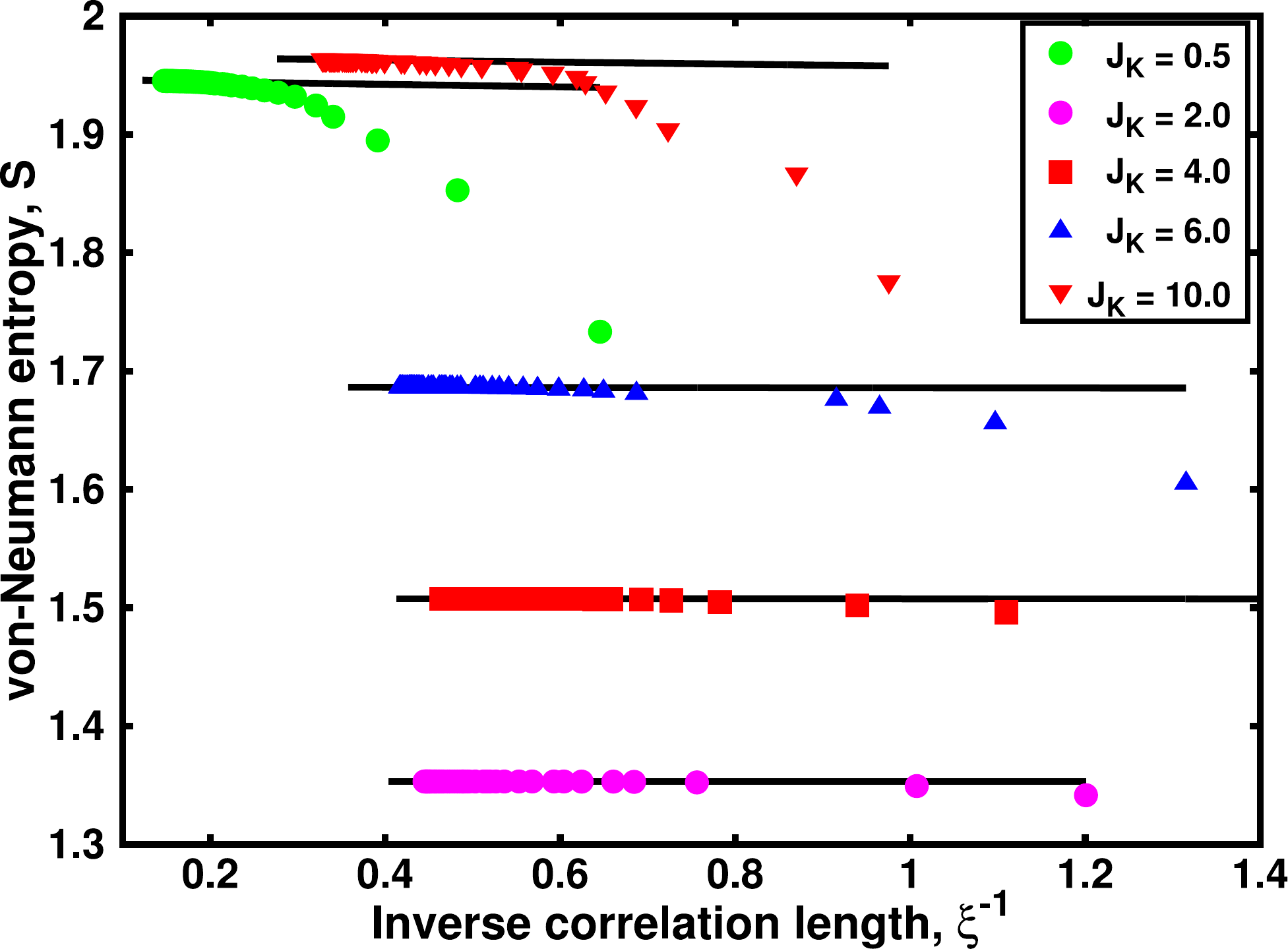}
  \caption{(Colour online) von-Neumann entropy $S$ versus inverse correlation length $\xi^{-1}$. Symbols are simulation data for different values of $J_K$ shown in the legend. Solid black lines are the fit of form $S = \alpha \ln \xi^{-1} + \beta$ where $\alpha$ and $\beta$ are fit parameters.}
  \label{fig:von-Neumann_entropy_vs_inverse_corr_len}
\end{figure}

The von-Neumann entropy $S$ in the main text are scaled with respect to the basis size $m$. Thus those values of $S$ corresponds to values of $S$ as $m \rightarrow \infty$ \cite{Nishino,Andersson,Tagliacozzo,Crosswhite,McCulloch}. This is done in a two-step procedure. First, the correlation length $\xi$ is scaled with respect to $m$ to obtain its value at $m \rightarrow \infty$ as shown in Fig. \ref{fig:inverse_corr_len_vs_m}. Since there are several correlation lengths corresponding to different quasi-particle excitations to choose from, the quasi-particle excitation that has the largest correlation length is chosen - in this case, the largest correlation length is that of the spin excitation. The different symbols in Fig. \ref{fig:inverse_corr_len_vs_m} are simulation data of $\xi^{-1}$ for values of $J_K$ shown in the legend. The black lines is the fit function of form $\xi^{-1} = a m^{-\kappa} + \xi_0^{-1}$, where $a$, $\kappa$ and $\xi_0^{-1}$ are fitting parameters. By fitting the fit function to the simulation data, $\xi_0^{-1}$ which is the $y$-axis intercept and represents the value of $\xi^{-1}$ at $m \rightarrow \infty$ is obtained.

Next, a function $S = \alpha \ln \xi^{-1} + \beta$ is fitted against the simulation data of $S$ and $\xi$. The fit function parameters here are $\alpha$ and $\beta$. These fitting parameters are then used together with $\xi_0^{-1}$ obtained from the previous procedure to obtain $S_0$ which is the value of $S$ as $m \rightarrow \infty$. This is shown in Fig. \ref{fig:von-Neumann_entropy_vs_inverse_corr_len}) where the different symbol are the simulation data for different $J_K$'s shown in the legend, and the black line is the fit function. This is the value of $S$ that is plotted the main text.

%%%%%%%%%%%%%%%%%%%%%%%%%%%%%%%%%%%%%%%%%%%%%%%%%%%%%%%%%%%%%%%%%%%%%%%%%%%%%%%%%%%%%%%%%%%%%%%%%%%%%%%%%%%

\subsection{Evaluation of the string order parameter $O_\text{string}^2$ in an iMPS}
\label{appendix_Ostring}
The string order parameter $O_\text{string}^2$ in Eq. (\ref{eqn:string_order}) of the main text defined as
\begin{eqnarray}
O_\text{string}^2 \equiv \lim_{|j-k| \rightarrow \infty} \left\langle  \mathbbm{1}_j \text{exp} \left[ \frac{i \pi}{2} \sum^k_{l=j} \left( \hat{n}_l - 1 \right) \right] \mathbbm{1}_k \right\rangle \nonumber
\end{eqnarray}
was used to determine 2-particle fluctuation in the region between sites $j$ and $k$ of the lattice. Since the exponent $\text{exp} \left[ \frac{i \pi}{2} \sum^k_{l=j} \left( \hat{n}_l - 1 \right) \right]$ does the same task as $(-1)^\frac{n_j - 1}{2}$, $O_\text{string}^2$ can be expressed as a correlation function \cite{Kolley}
\begin{eqnarray}
O_\text{string}^2 = \lim_{|j-k| \rightarrow \infty} \braket{p(j) p(k)}
\label{eqn:string_order2}
\end{eqnarray}
where $p(j) = \prod_{i < j} (-1)^\frac{n_i - 1}{2}$ is the ``kink" operator and
\begin{eqnarray}
\braket{p(j) p(k)} = \left\langle \prod_{i < j} \mathbbm{1}_i \prod_{i=j}^k (-1)^\frac{n_i - 1}{2} \prod_{i>k} \mathbbm{1}_i \right\rangle
\label{eqn:string_order3}
\end{eqnarray}
which yields the same result as Eq. (\ref{eqn:string_order}). This makes $O^2_\text{string}$ appear similar to a local order parameter, e.g. $m^2 = \braket{M^2}$ where $M$ is the magnetization.

In calculating $O_\text{string}^2$ in an iMPS, an extensive order parameter $P = \sum_i p(i)$ is initially constructed from the kink operators. Since the sign of $\braket{P}$ is indeterminate, it cannot be directly evaluated, however, $\braket{P^2}$ is always positive and it is this value that is related to $O_\text{string}^2$ via
\begin{eqnarray}
O_\text{string}^2 = \frac{\braket{P^2}}{L^2}
\end{eqnarray} 
where $L = b \xi$, $\xi$ is the correlation length and $b$ is a scaling factor. This is equivalent to calculating the $O_\text{string}^2$ over a finite section of size $L = b\xi$ of the infinite lattice. The expectation value of an $n$th power of an operator $P^n$ in an iMPS is obtained as a degree $n$ polynomial of the lattice size $L$, which is exact in the asymptotic large-$L$ limit. Hence, $O_\text{string}^2 = \braket{P^2}/L^2$ is evaluated directly as the coefficient of the degree 2 component of $\braket{P^2}$.

%%%%%%%%%%%%%%%%%%%%%%%%%%%%%%%%%%%%%%%%%%%%%%%%%%%%%%%%%%%%%%%%%%%%%%%%%%%%%%%%%%%%%%%%%%%%%%%%%%%%%%%%%%%

\subsection{von-Neumann entropy $S$ in the large $U$ limit}
\label{appendix_largeU}
\begin{figure}[h!]
  \centering\includegraphics[width=0.45\textwidth]{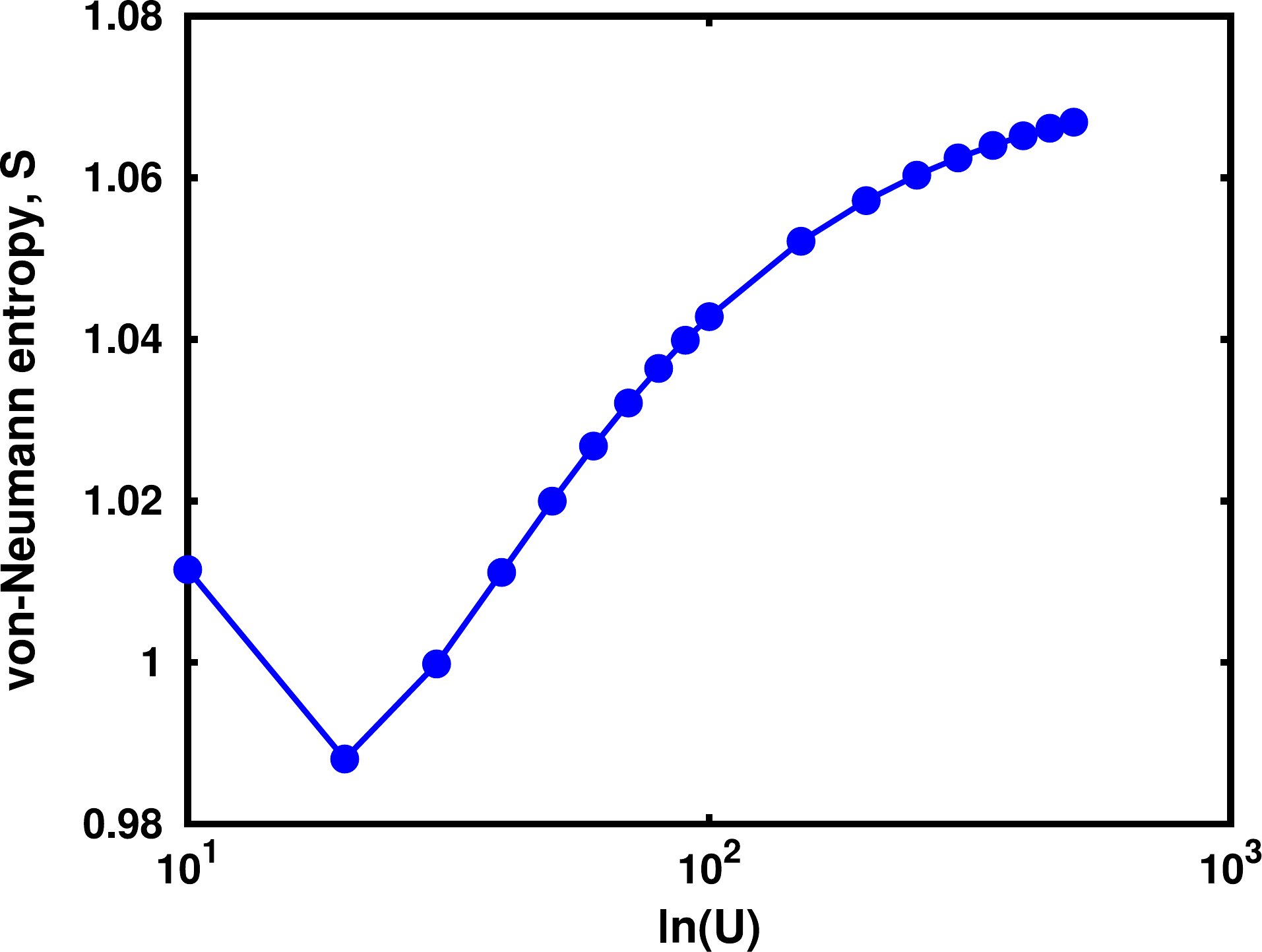}
  \caption{von-Neumann entropy $S$ versus $\ln(U)$, where $U$ is the Hubbard interaction strength. Other parameters are $J_K = 2$ and $J_\perp = 0$. This plot is an extension of Fig. \ref{fig:svn-U-J_K} when $U > 20$. The gradual increase of $S$ even when $U$ is large suggests that $S$ is lower bounded.}
  \label{fig:svn-U-J_K-large_U}
\end{figure}

\begin{figure}[h!]
  \centering\includegraphics[width=0.45\textwidth]{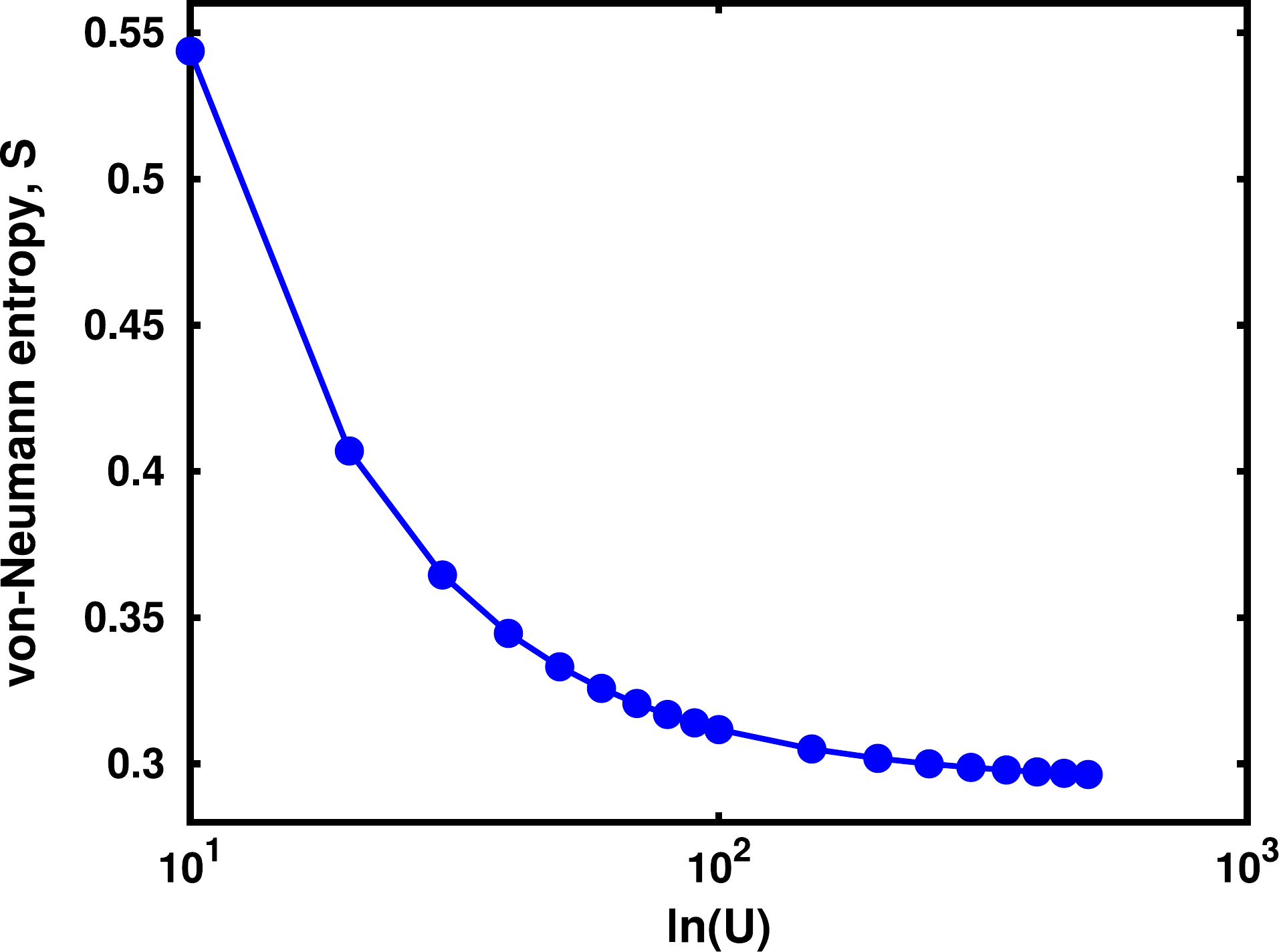}
  \caption{von-Neumann entropy $S$ versus $\ln(U)$, where $U$ is the Hubbard interaction strength. Other parameters are $J_K = 0$ and $J_\perp = 1$. This plot is an extension of Fig. \ref{fig:svn-U-Jperp} when $U > 20$. The rapid decay of $S$ as $U$ is increased indicates that $S \rightarrow 0$ as $U \rightarrow \infty$.}
  \label{fig:svn-U-Jperp-large_U}
\end{figure}

This section shows the von-Neumann entropy $S$ versus Hubbard interaction $U$ when the latter is unphysically large. The purpose of providing these graphs are to support the claims made in Section \ref{Coulomb-svn1} regarding the behavior of $S$ when $U > 20$ for the SPT phase and the topologically trivial phase. In the main text, the claim was made that in the SPT phase, $S$ was lower bounded even though $U$ was increased. This is evident in Fig. \ref{fig:svn-U-J_K-large_U} where there is no sign of a decreasing $S$ even though $U \sim \mathcal{O}(10^2)$. In fact, the minimum of $S$ in this plot is at $U = 20$, which is the largest value of $U$ shown in Fig. \ref{fig:svn-U-J_K} the main text.

Fig. \ref{fig:svn-U-Jperp-large_U} is an extension of Fig. \ref{fig:svn-U-Jperp}. In the topologically trivial case, $S$ in Section \ref{Coulomb-svn1} was claimed to have no lower bound. This can be seen in Fig. \ref{fig:svn-U-Jperp-large_U} where $S$ continuous to decay as $U$ is increased.

%%%%%%%%%%%%%%%%%%%%%%%%%%%%%%%%%%%%%%%%%%%%%%%%%%%%%%%%%%%%%%%%%%%%%%%%%%%%%%%%%%%%%%%%%%%%%%%%%%%%%%%%%%%
%%%%%%%%%%%%%%%%%%%%%%%%%%%%%%%%%%%%%%%%%%%%%%%%%%%%%%%%%%%%%%%%%%%%%%%%%%%%%%%%%%%%%%%%%%%%%%%%%%%%%%%%%%%

\end{document}